\documentclass{article}

\PassOptionsToPackage{numbers, compress}{natbib}
\usepackage[main,final]{neurips_2026}

\usepackage[utf8]{inputenc} 
\usepackage[T1]{fontenc}    
\usepackage{hyperref}       
\hypersetup{hidelinks}
\usepackage{url}            
\usepackage{booktabs}       
\usepackage{amsfonts}       
\usepackage{nicefrac}       
\usepackage{microtype}      
\usepackage[table]{xcolor}  
\usepackage{graphicx}
\usepackage{caption}
\usepackage{float}
\usepackage{algorithm}     
\usepackage{algorithmic}   
\usepackage{amsmath}
\usepackage{tabularx}
\usepackage{threeparttable}
\usepackage{enumitem}
\usepackage{multirow}
\usepackage{etoolbox}
\usepackage{pgfplots}
\usepgfplotslibrary{groupplots}
\pgfplotsset{compat=1.18}

\AtBeginEnvironment{table}{\tiny}
\AtBeginEnvironment{table*}{\tiny}
\AtBeginEnvironment{table}{\setlength{\tabcolsep}{3pt}}
\AtBeginEnvironment{table*}{\setlength{\tabcolsep}{3pt}}

\title{Discovering Programmatic Policies from Reinforcement Learning-Based Traffic Signal Controllers}

\author{%
  \textbf{Lindong Xie}$^{1,2}$\hspace{0.5em}\textbf{Yang Zhang}$^{1}$\hspace{0.5em}\textbf{Beiyu Song}$^{1}$\hspace{0.5em}\textbf{Zeren Xing}$^{1}$\hspace{0.5em}\textbf{Flora D. Salim}$^{2}$\hspace{0.5em}\textbf{Edward Chung}$^{1}$\thanks{Corresponding author.} \\[7pt]
  $^{1}$Department of Electrical and Electronic Engineering \\
  The Hong Kong Polytechnic University, Hong Kong, China \\
  $^{2}$School of Computer Science and Engineering \\
  University of New South Wales, Sydney, Australia \\
  \texttt{lindong.xie@connect.polyu.hk}\quad\texttt{edward.cs.chung@polyu.edu.hk}
}

\begin{document}

\maketitle

\begin{abstract}

    Traffic signal control (TSC) is critical for mitigating urban congestion. Deep reinforcement learning (DRL) has achieved strong performance by learning adaptive policies from traffic observations, but learned neural controllers are often opaque, structurally complex, and difficult to generalize across scenarios. We propose PPL-TSC, a programmatic policy learning framework that distills compact and human-readable TSC policies from DRL teachers. PPL-TSC formulates policy learning as an evolutionary search over phase utility functions (PUFs), which score candidate signal phases and select the highest-utility phase for execution. Each generation consists of four steps: policy generation, policy review, policy calibration, and policy evaluation. In the generation step, a large language model (LLM) acts as a semantics-aware evolutionary operator to propose and refine candidate PUFs. Generated PUFs are screened through explicit checks for syntactic validity, mathematical soundness, and consistency with TSC principles, with invalid ones repaired through iterative feedback. Valid PUFs are then calibrated to improve behavioral fidelity to the DRL teacher, and evaluated using a joint score combining teacher fidelity and traffic-control performance to select candidates for the next generation. Extensive experiments across diverse traffic scenarios and representative DRL-based controllers show that PPL-TSC produces compact and human-readable policies, achieves competitive performance for most teacher families, improves transfer generalization, and substantially reduces computational cost. Code is available at \url{https://github.com/ForrestXie9/PPL-TSC}.

\end{abstract}

\section{Introduction}
Traffic signal control (TSC) is a core problem in intelligent transportation systems, with direct implications for urban congestion, travel efficiency, and transportation-related emissions \citep{wei2019survey,wu2025big}. Classical approaches to TSC include fixed-time control \citep{webster1958traffic,little1966synchronization,little1981maxband}, which relies on predetermined timing plans derived from historical demand, and adaptive control \citep{cools2007self,varaiya2013max,hunt1981scoot,lowrie1990scats}, which adjusts signal decisions based on real-time traffic observations. While these paradigms have supported decades of deployment, they either lack responsiveness to non-stationary traffic patterns or depend heavily on manually designed control rules and simplified modeling assumptions (e.g., stationary demand) \citep{wei2021recent}. Designing effective TSC policies that are both adaptive and practically deployable remains a fundamental challenge.

\begin{figure*}[t]
    \centering 
    \begin{minipage}[t]{0.51\textwidth} 
        \vspace{0pt} 
	        \captionof{table}{Transfer performance of DRL-based TSC methods under Manhattan traffic-flow shifts and Manhattan-to-Los Angeles transfer.}
	        \label{tab:intro_shift_challenge}
	        \renewcommand{\arraystretch}{1.3}
	        \vspace{0.em} 
	        \footnotesize
	        \setlength{\tabcolsep}{4pt}
	        \resizebox{0.98\linewidth}{!}{
            \begin{tabular}{l cc cc}
              \toprule
              \multicolumn{1}{c}{\multirow{2}{*}{\textbf{Teacher}}} & \multicolumn{2}{c}{$\Delta$ATT (\%) $\uparrow$} & \multicolumn{2}{c}{Win Rate (\%) $\uparrow$} \\
              \cmidrule(lr){2-3} \cmidrule(lr){4-5}
               & In-dom. & Cross-scen. & In-dom. & Cross-scen. \\
            \midrule
            CoLight~\citep{wei2019colight} & $-0.24 \pm 6.47$ & N/A & 50.0\% & N/A \\
            EcoLight~\citep{chauhan2020ecolight} & $-0.37 \pm 6.60$ & $-51.79 \pm 42.43$ & 50.0\% & 8.0\% \\
            FRAP~\citep{zheng2019learning} & $-0.07 \pm 3.60$ & $-20.73 \pm 38.45$ & 65.0\% & 24.0\% \\
            MPLight~\citep{chen2020toward} & $-0.14 \pm 5.60$ & $-2.34 \pm 32.67$ & 45.0\% & 28.0\% \\
              TinyLight~\citep{xing2022tinylight} & $-62.62 \pm 172.58$ & N/A & 50.0\% & N/A \\
	            \bottomrule
	          \end{tabular}
	        }
	        \vspace{0.em} 
        {\scriptsize \textit{Note:} ATT denotes average travel time. $\Delta$ATT denotes relative ATT improvement under transfer; negative values mean the transferred controller performs worse (i.e., ATT increases). ATT win rate is the fraction of transfer cases in which ATT does not increase. N/A indicates methods not applicable in cross-scenario transfer.\par}
    \end{minipage}
    \hfill 
    \begin{minipage}[t]{0.44\textwidth}
        \vspace{0pt} 
        \centering
        \includegraphics[width=\linewidth]{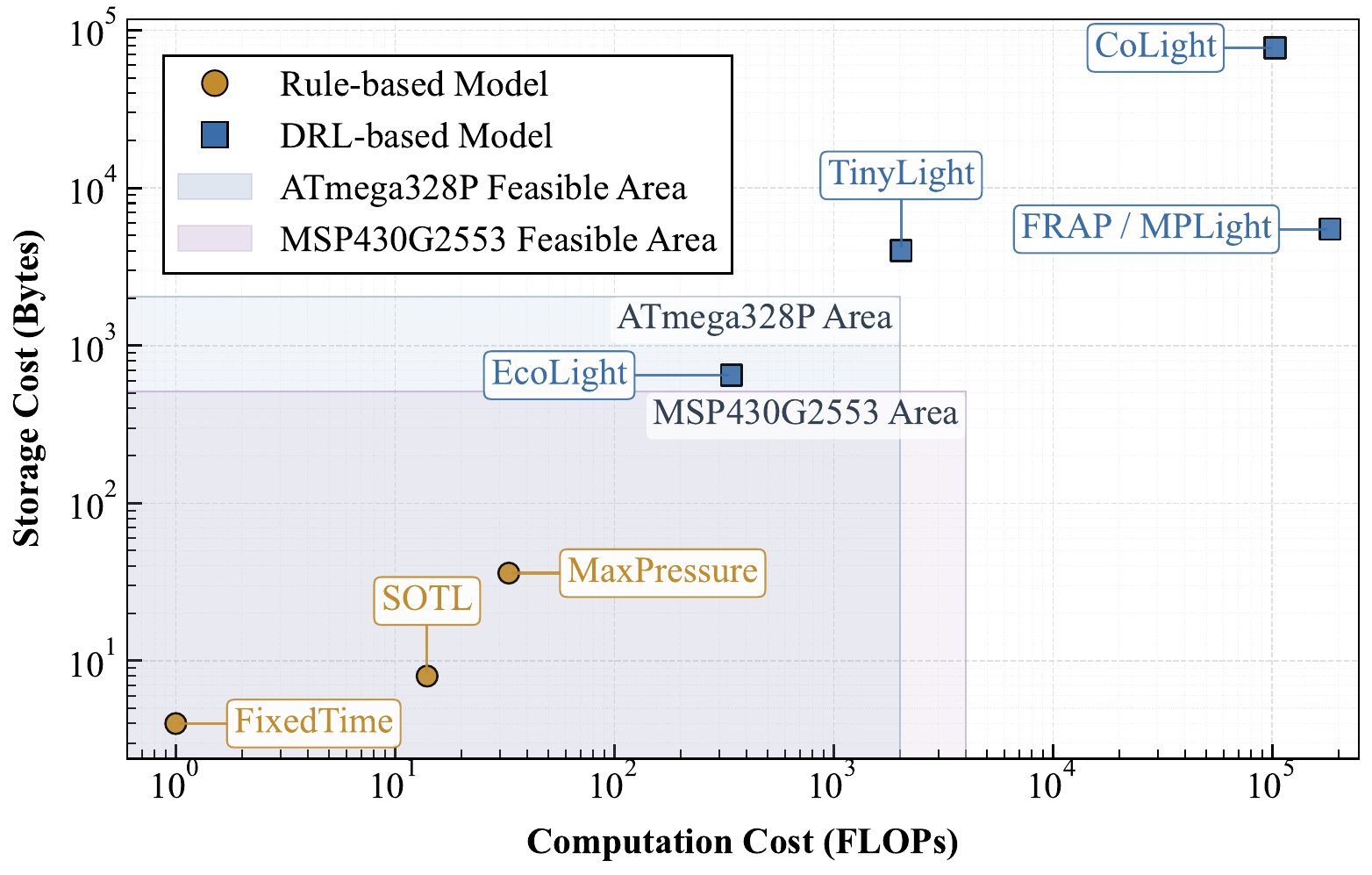}
  
        \captionof{figure}{Storage and computational costs of DRL-based and rule-based TSC methods.}
        \label{fig:intro_resource_challenge}
    \end{minipage}
    \vspace{-15pt} 
\end{figure*}

Deep reinforcement learning (DRL) has become a prominent paradigm for adaptive TSC \citep{li2016traffic,yau2017survey,xiao2025advances}. Through simulator interaction, DRL agents learn neural policies that map traffic states to signal phase decisions, achieving strong empirical performance across diverse traffic networks \citep{wei2019colight,zheng2019learning}. Despite this progress, deploying neural controllers in practice remains challenging. First, learned policies lack explicit traffic-control concepts (e.g., queue pressure, downstream congestion), making it difficult to inspect whether their decisions follow established traffic-control principles \citep{zhu2022extracting}.
Second, policies trained for specific traffic patterns often degrade under distribution shifts \citep{gu2024pi}. As shown in Table~\ref{tab:intro_shift_challenge}, most representative DRL methods remain stable under in-domain traffic-flow variations in Manhattan but show performance degradation when transferred to Los Angeles. Under cross-scenario transfer, methods like EcoLight~\citep{chauhan2020ecolight}, FRAP~\citep{zheng2019learning}, and MPLight~\citep{chen2020toward} suffer performance drops, while others (e.g., CoLight~\citep{wei2019colight}, TinyLight~\citep{xing2022tinylight}) are inapplicable due to action-space mismatches (e.g., differing phase definitions and dimensionality), preventing direct policy reuse without architectural adaptation. Addressing such shifts typically requires retraining with additional data and computation.
Third, neural controllers impose nontrivial memory and computational costs during real-time inference \citep{xing2022tinylight}. Figure~\ref{fig:intro_resource_challenge} compares storage and floating-point operations (FLOPs) per decision. DRL-based methods require substantially more resources than classical rule-based controllers (e.g., fixed-time \citep{webster1958traffic}, SOTL \citep{cools2007self}, and MaxPressure \citep{varaiya2013max}). This resource gap is particularly consequential for deployment on low-resource edge platforms.

These limitations have motivated growing interest in explicit policy learning for TSC, where controllers are represented as rules or programs rather than black-box neural networks \citep{gu2024pi,liao2025gplight+,liao2025symlight}. Existing methods construct such policies using Monte Carlo tree search \citep{swiechowski2023monte} or genetic programming \citep{koza1990genetic}, but they typically search from scratch through extensive interaction with the traffic environment. This process is costly in large combinatorial policy spaces and fails to leverage the control knowledge already encoded in DRL teachers. Moreover, these methods provide limited guarantees that the discovered rules remain semantically consistent with established TSC principles.

To address these challenges, we propose PPL-TSC, a unified programmatic policy learning framework that learns explicit and human-readable TSC policies from diverse DRL teachers. Instead of discovering explicit policies through trial-and-error exploration from scratch, PPL-TSC leverages rollout data collected from a pretrained DRL teacher to guide policy learning. We represent the student policy as a phase utility function (PUF), an explicit phase-scoring function that scores candidate phases using phase-level traffic features and selects the highest-scoring phase. Its explicit expression makes the contribution of each term inspectable. Overall, our main contributions are summarized as follows:



\begin{itemize}[leftmargin=*, itemsep=0pt, topsep=2pt]

\item We propose PPL-TSC, a teacher-guided programmatic policy learning framework for TSC that distills explicit policies from pretrained DRL teachers, reducing reliance on costly policy search from scratch through extensive environment interaction.

\item We formulate policy learning as an iterative generate--review--calibrate--evaluate process over phase utility functions. The process integrates LLM-based program search, explicit validity checks, intersection-specific calibration, and joint evaluation to learn human-readable policies aligned with teacher behavior and TSC principles.

\item We conduct extensive experiments across representative DRL-based TSC methods and diverse traffic scenarios. PPL-TSC learns human-readable student policies that achieve competitive traffic performance, improve generalization under transfer settings, and substantially reduce storage and computational costs relative to their DRL teachers.


\end{itemize}



\section{Related Work}
Early TSC systems relied mainly on fixed-time strategies, such as Webster's method \citep{webster1958traffic}, GreenWave \citep{little1966synchronization}, and MAXBAND \citep{little1981maxband}, where signal timings are derived from historical demand and fixed during deployment. Adaptive methods later introduced online updates based on real-time traffic observations, including rule-based heuristics such as SOTL \citep{cools2007self} and MaxPressure \citep{varaiya2013max}, and model-based systems such as SCOOT \citep{hunt1981scoot} and SCATS \citep{lowrie1990scats}. More recently, DRL has become a prominent paradigm for learning adaptive policies directly from traffic observations, reducing reliance on hand-designed rules and simplified traffic assumptions.

Existing DRL-based TSC methods can be broadly divided into value-based and policy-based approaches. Value-based methods, including FRAP \citep{zheng2019learning}, MPLight \citep{chen2020toward}, CoLight \citep{wei2019colight}, and TinyLight \citep{xing2022tinylight}, estimate the long-term value of each candidate phase and select the phase with the highest value. Policy-based methods, such as A-CATs \citep{aslani2017adaptive}, policy-gradient methods \citep{rizzo2019time}, multi-agent actor--critic methods \citep{chu2019multi}, and AttendLight \citep{oroojlooy2020attendlight}, directly learn mappings from traffic states to control actions. Despite their empirical success, these neural controllers remain difficult to interpret, validate, and deploy in resource-constrained settings \citep{gu2024pi}. These limitations have motivated interest in explicit policy representations for TSC. Early work considered predefined polynomial control forms \citep{ault2019learning}, whose parameters were fitted to mimic trained neural controllers. However, such fixed structures are often too restrictive to capture complex decision logic. More recent methods, such as $\pi$-Light \citep{gu2024pi}, GPLight+ \citep{liao2025gplight+}, and SymLight \citep{liao2025symlight}, use program search operators, including Monte Carlo tree search \citep{swiechowski2023monte} and genetic programming \citep{ahvanooey2019survey}, to construct explicit control policies through interaction with the traffic environment. While these methods improve transparency, they typically search from scratch in a large combinatorial policy space, require substantial environment interaction, and make limited use of the knowledge already encoded in strong DRL teachers. In addition, evaluating candidate policies mainly through traffic-control performance does not guarantee that the resulting expressions are mathematically valid or semantically meaningful.

PPL-TSC differs from these approaches by learning programmatic policies directly from pretrained DRL teachers rather than discovering them from scratch. It uses teacher rollout data to guide policy learning and integrates generation, review, calibration, and evaluation, with explicit checks of generated functions and joint evaluation of teacher fidelity and control performance.

\section{Method}
We first formalize teacher-guided policy learning for TSC and define the student policy representation. We then describe the four steps of PPL-TSC: generation, review, calibration, and evaluation.
\subsection{Problem Formulation}
We consider a traffic network with $M$ signalized intersections. At decision step $t$, intersection $i$ observes a local traffic state $\mathbf{s}_t^{(i)}$ and selects a signal phase $p_t^{(i)}$ from a feasible set $\mathcal{P}_t^{(i)}$. A signal phase is a predefined set of compatible, non-conflicting traffic movements that receive the right-of-way concurrently. This feasible set depends on the controller's action space: phase-selection controllers directly choose among all feasible phases, whereas keep/switch controllers choose between keeping the current phase and switching to the next one, i.e., $\mathcal{P}_t^{(i)}=\{p_{t,\mathrm{cur}}^{(i)}, \mathrm{next}(p_{t,\mathrm{cur}}^{(i)})\}$. Definitions of TSC terms are provided in Appendix~\ref{app:definitions}.

In DRL-based TSC, a neural policy $\pi^{\mathrm{Tch}}$, referred to as the teacher policy, maps traffic states to signal phases and is learned through interaction with the environment. Executing the teacher over the network produces rollout data $\mathcal{D}=\bigcup_{i=1}^{M}\mathcal{D}_i$, where $\mathcal{D}_i=\{(\mathbf{s}_t^{(i)},p_t^{\mathrm{Tch},(i)})\}_{t=1}^{N_i}$ contains the observed state and teacher-selected phase for intersection $i$. Background on DRL-based TSC is provided in Appendix~\ref{app:drl_background}.

Our objective is to learn a compact and human-readable student policy $\pi^{\mathrm{Stu}}$ that maps the same traffic states to phase decisions, approximates the teacher's behavior, and maintains competitive traffic-control performance.

\subsection{Student Policy Representation}

We represent the student policy as a phase utility function (PUF) that scores candidate phases. Let $\mathbf{x}_{p,t}^{(i)}$ denote the traffic features used to score candidate phase $p \in \mathcal{P}_t^{(i)}$, constructed from the local state $\mathbf{s}_t^{(i)}$ according to the teacher's input and action interfaces. The utility of phase $p$ is defined as
\begin{equation}
U_{p,t}^{(i)} = f(\mathbf{x}_{p,t}^{(i)}) + \beta_{p}^{(i)},
\label{eq:puf}
\end{equation}
where $f(\cdot)$ is a programmatic expression shared across intersections to capture common decision logic, and $\beta_{p}^{(i)}$ is an intersection- and phase-specific bias term for local adaptation. For direct cross-scenario transfer, we retain $f(\cdot)$ and set $\beta_p^{(i)}=0$ at each target intersection. For target-specific adaptation, the phase biases can be fine-tuned while keeping $f(\cdot)$ fixed.

The student policy then selects the phase with the highest utility:
\begin{equation}
p_t^{\mathrm{Stu}, (i)} = \arg\max_{p \in \mathcal{P}_t^{(i)}} U_{p,t}^{(i)}.
\label{eq:action}
\end{equation}

\subsection{PPL-TSC Framework}
\label{sec:framework}

\begin{figure*}[t]
\centering
\includegraphics[width=0.9\linewidth]{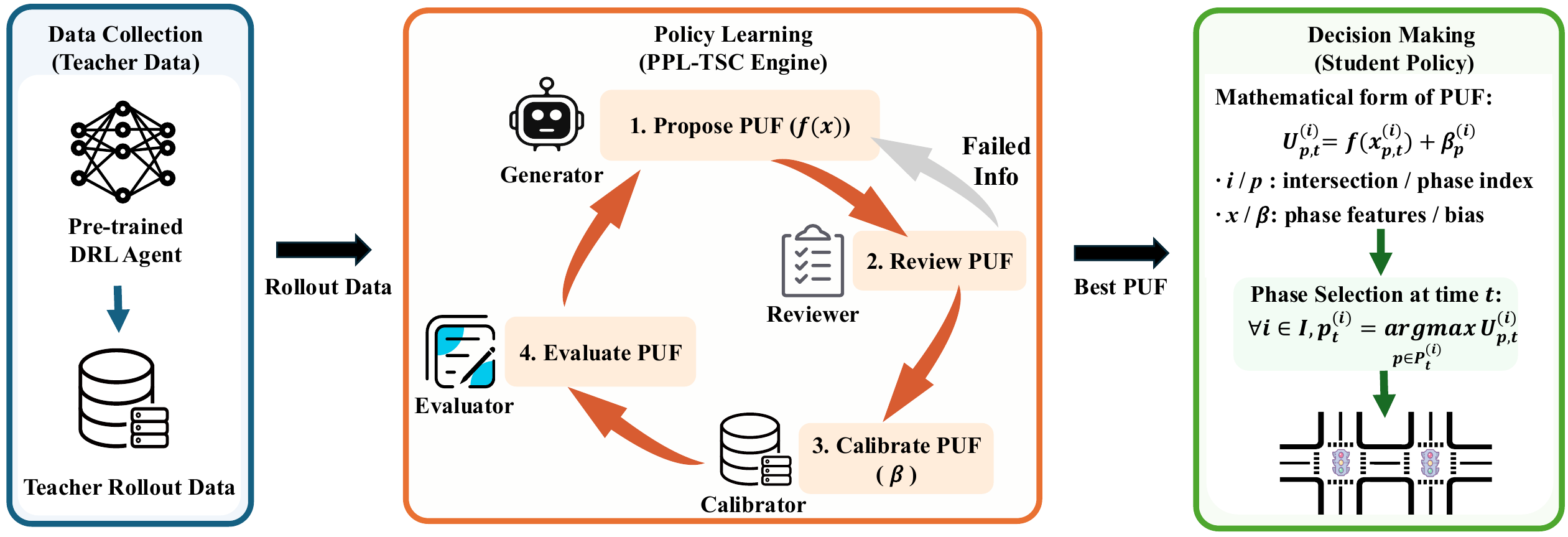}
\caption{Overview of PPL-TSC. The framework consists of three modules: data collection, policy learning, and decision making. Data collection gathers rollout samples from a pretrained DRL-based TSC teacher. Policy learning obtains a phase utility function (PUF) through iterative generation, review, calibration, and evaluation. Decision making uses the learned PUF $f(\mathbf{x})$ with intersection-specific phase bias terms $\boldsymbol{\beta}$ for real-time signal phase selection.}
\label{fig:Workflow of PPL-TSC}
\end{figure*}

PPL-TSC searches for the PUF in Eq.~\eqref{eq:puf} through an evolutionary learning process with four steps: generation, review, calibration, and evaluation, as illustrated in Fig.~\ref{fig:Workflow of PPL-TSC}. The search proceeds for up to $G_{\mathrm{max}}$ generations with a population size $N_{\mathrm{pop}}$.

\subsubsection{Policy Generation}

The generation module uses an LLM as a program synthesis operator to propose candidate phase-scoring functions $f(\cdot)$ from the teacher rollout data $\mathcal{D}$ \citep{xie2025co,liu2024evolution}. Its goal is to explore the space of readable phase-scoring rules while keeping generated candidates consistent with the available traffic features and basic TSC principles.

We use two prompt types. The initialization prompt $\mathcal{T}(\text{Task}_{\text{init}}, \text{Ctx}_{\text{init}}, \text{Cons}, \text{Out})$ creates the initial population from feature descriptions derived from teacher states. The evolution prompt $\mathcal{T}(\text{Task}_{\text{evol}}, \text{Ctx}_{\text{evol}}, \text{Cons}, \text{Out})$ generates offspring by conditioning on parent PUFs and their evaluation feedback. In both cases, $\text{Cons}$ specifies syntactic, mathematical, and semantic requirements: generated code must be executable, numerical operations must remain robust under extreme inputs, and feature use should respect TSC principles such as monotonicity with respect to traffic load. The output specification (i.e., $\text{Out}$) requires a candidate PUF together with a brief design rationale. Detailed example prompts and constraints are provided in Appendices~\ref{app:prompt_templates} and~\ref{app:constraints}.

Initialization prompts are used until the population reaches $N_{\mathrm{pop}}$, after which evolution prompts generate offspring from selected parent PUFs.

\subsubsection{Policy Review}

After receiving candidate PUFs from the generation step, the policy review module checks whether the generated functions satisfy syntactic correctness, mathematical validity, and semantic consistency under the specified constraints. The review checks the generated function $f$ before calibration; the calibrated biases then adjust the scores of individual phases. This explicit verification is needed because LLM-generated functions may still violate constraints despite prompt guidance \citep{joel2024survey}. Candidates that fail verification are marked as invalid and returned to the LLM with diagnostic feedback for revision. An example repair prompt is provided in Appendix~\ref{app:prompt_templates}.

\subsubsection{Policy Calibration}
Policy calibration adjusts the bias terms in Eq.~\eqref{eq:puf} after a candidate PUF passes review. The phase-scoring function $f(\cdot)$ is kept fixed and shared across intersections, while the phase biases are learned separately for each intersection from teacher rollouts. This design improves the student's agreement with local teacher decisions while preserving the generated rule and avoiding further interaction with the traffic environment.

For intersection $i$, let $\mathcal{P}^{(i)}=\bigcup_{t=1}^{N_i}\mathcal{P}_t^{(i)}$ denote the union of candidate phase sets over the rollouts. We assign one bias term $\beta_p^{(i)}$ to each phase $p\in\mathcal{P}^{(i)}$, collected in $\boldsymbol{\beta}^{(i)}$. We optimize $\boldsymbol{\beta}^{(i)}$ to maximize agreement with teacher decisions on $\mathcal{D}_i$:
\begin{equation}
\max_{\boldsymbol{\beta}^{(i)}} \frac{1}{N_i} \sum_{t=1}^{N_i} \mathbf{1} \left( p_t^{\mathrm{Stu},(i)} = p_t^{\mathrm{Tch},(i)} \right),
\label{eq:calibration}
\end{equation}
where $\mathbf{1}(\cdot)$ is the indicator function, which equals 1 when the student and teacher select the same phase and 0 otherwise. Since this objective function is non-differentiable, we use CMA-ES \citep{hansen2003reducing}, a robust black-box optimization algorithm, to optimize $\boldsymbol{\beta}^{(i)}$ independently for each intersection.

\subsubsection{Policy Evaluation}
Policy evaluation ranks calibrated PUFs according to both teacher fidelity and closed-loop traffic performance. Fidelity measures whether the student reproduces the teacher's phase choices, while traffic performance assesses whether the student remains effective when deployed in the traffic environment. We combine the two criteria using the joint score
\begin{equation}
J(U) = w_1 F(U) + w_2 P(U),
\label{eq:evaluation}
\end{equation}
where $F(U)$ denotes behavioral fidelity, $P(U)$ denotes traffic performance relative to the teacher, and $w_1,w_2\geq0$ balance the two terms. In our experiments, we use equal weights, $w_1=w_2=1$, for teacher-policy fidelity and downstream control performance.

\noindent\textbf{Behavioral Fidelity $F(U)$.} We measure fidelity on the teacher rollout dataset $\mathcal{D}$ used in calibration, with student decisions given by Eq.~\eqref{eq:action}. The fidelity score is the fraction of rollout samples on which the student and teacher select the same phase:
\begin{equation}
F(U) = \frac{1}{|\mathcal{D}|} \sum_{i=1}^{M} \sum_{t=1}^{N_i} \mathbf{1}\left( p_t^{\mathrm{Stu},(i)} = p_t^{\mathrm{Tch},(i)} \right),
\label{eq:agreement}
\end{equation}
where $\mathbf{1}(\cdot)$ is the indicator function and $|\mathcal{D}| = \sum_{i=1}^{M} N_i$. This metric lies in $[0,1]$, with $1$ indicating perfect agreement with the teacher on the rollout data.

\noindent\textbf{Control Performance $P(U)$.} We evaluate traffic control performance by deploying the student policy in the same TSC environment used to test the teacher. Let $ATT^{\mathrm{Stu}}$ and $ATT^{\mathrm{Tch}}$ denote the average travel times of the student and teacher controllers, respectively. The relative improvement over the teacher is
\begin{equation}
\Delta = \frac{ATT^{\mathrm{Tch}} - ATT^{\mathrm{Stu}}}{|ATT^{\mathrm{Tch}}| + \epsilon},
\label{eq:delta}
\end{equation}
where $\epsilon = 10^{-4}$ is used for numerical stability. A positive $\Delta$ indicates that the student achieves a lower average travel time than the teacher, $\Delta=0$ indicates equal performance, and $\Delta<0$ indicates worse performance.

We convert this improvement to a bounded performance score in $(0,1]$ as
\begin{equation}
P(U) = \exp\left(\min(\Delta, 0)\right).
\label{eq:performance}
\end{equation}
This score equals $1$ when the student matches or outperforms the teacher and decreases smoothly when the student performs worse.

After evaluation, each offspring is compared with its parent using the joint score in Eq.~\eqref{eq:evaluation}. The offspring replaces its parent only if it achieves a higher score; otherwise, the parent is retained. The search terminates when the maximum number of generations $G_{\mathrm{max}}$ is reached.
\section{Experiments}
We evaluate PPL-TSC through six research questions.
\textbf{RQ1 (Performance):} Can PPL-TSC match the traffic-control performance of its DRL teacher under identical evaluation settings?
\textbf{RQ2 (Fidelity):} How accurately does PPL-TSC reproduce the teacher's decision behavior?
\textbf{RQ3 (Generalizability):} Does PPL-TSC generalize under same-scenario different-traffic-flow and cross-scenario settings?
\textbf{RQ4 (Readability):} Can the learned student be expressed as a compact policy consistent with traffic control principles?
\textbf{RQ5 (Resource Efficiency):} Is PPL-TSC more resource-efficient than neural teachers?
\textbf{RQ6 (Ablation):} How do policy generation, calibration, and evaluation affect the effectiveness of PPL-TSC?

\subsection{Experimental Setup}

\paragraph{Simulator and traffic scenarios.}
We conduct experiments in CityFlow \citep{zhang2019cityflow}, an open-source microscopic traffic simulator. The benchmark includes single-intersection, corridor, and grid scenarios from seven road networks derived from five real-world urban areas \citep{mei2024libsignal}: three $1\times1$ Hangzhou networks (Hangzhou1/2/3), a $5\times1$ Atlanta corridor, a $3\times4$ Jinan grid, a $4\times4$ Los Angeles grid, and a $16\times3$ Manhattan grid. Jinan and Manhattan are structurally homogeneous, with all non-virtual signalized intersections sharing the same phase and link configurations, whereas Atlanta and Los Angeles are structurally heterogeneous due to multiple intersection templates. Each road network defines a scenario. For each scenario, we use one real-world traffic flow and generate four additional flows by perturbing each vehicle's departure time with uniform noise in $[-60\text{s},60\text{s}]$.

\paragraph{Teacher controllers and metrics.}
We use five DRL-based TSC methods as teacher controllers: CoLight \citep{wei2019colight}, FRAP \citep{zheng2019learning}, MPLight \citep{chen2020toward}, EcoLight \citep{chauhan2020ecolight}, and TinyLight \citep{xing2022tinylight}. EcoLight natively uses a keep/switch action interface, so we report EcoLight (native) in this original two-action setting and additionally evaluate an EcoLight (phase-selection) variant in which the student directly scores all feasible phases, consistent with the other teacher families. We evaluate PPL-TSC using average travel time (ATT) for control performance and average behavioral fidelity (ABF) for teacher-policy fidelity. We also report relative ATT improvement ($\Delta$ATT), defined as the student's relative ATT reduction over the teacher, and ATT wins, where the student achieves ATT no worse than its teacher.

  \subsection{Performance and Fidelity (RQ1--RQ2)}
  We first evaluate whether PPL-TSC preserves the control performance and decision behavior of its DRL teachers in the original deployment setting. PPL-TSC (Tailored) learns a separate policy for each teacher--scenario--flow combination, whereas PPL-TSC (Unified) learns a shared programmatic policy for each teacher family across multiple scenarios. Table~\ref{tab:main_results} reports $\Delta$ATT, ATT wins, and ABF over 35 scenario--flow combinations for each teacher family.
  \begin{table*}[t]
    \centering
    \caption{Performance and fidelity of PPL-TSC over 35 scenario--flow combinations. }
    \label{tab:main_results}
    \scriptsize
    \begin{threeparttable}
      \setlength{\tabcolsep}{5pt}
      \begin{tabular}{lcccccc}
        \toprule
        \multirow{2}{*}{\textbf{Teacher}} &
        \multicolumn{3}{c}{\textbf{PPL-TSC (Tailored)}} &
        \multicolumn{3}{c}{\textbf{PPL-TSC (Unified)}} \\
        \cmidrule(lr){2-4}\cmidrule(lr){5-7}
        & $\Delta$ATT (\%) $\uparrow$ & ATT wins $\uparrow$ & ABF $\uparrow$
        & $\Delta$ATT (\%) $\uparrow$ & ATT wins $\uparrow$ & ABF $\uparrow$ \\
        \midrule
        CoLight & $\boldsymbol{11.51 \pm 23.98}$ & 17/35 & $0.71 \pm 0.17$ & $\boldsymbol{16.70 \pm 32.98}$ & 18/35 & $0.60 \pm 0.22$ \\
        FRAP & $\boldsymbol{11.79 \pm 27.12}$ & 24/35 & $0.63 \pm 0.16$ & $\boldsymbol{12.07 \pm 32.99}$ & 22/35 & $0.58 \pm 0.19$ \\
	        MPLight & $\boldsymbol{0.54 \pm 6.16}$ & 21/35 & $0.73 \pm 0.12$ & $-0.03 \pm 10.96$ & 17/35 & $0.69 \pm 0.13$ \\
	        EcoLight (native) & $-18.05 \pm 41.66$ & 15/35 & $0.84 \pm 0.07$ & $-23.27 \pm 36.17$ & 8/35 & $0.82 \pm 0.07$ \\
	        EcoLight (phase-selection) & $\boldsymbol{22.70 \pm 27.23}$ & 30/35 & $0.48 \pm 0.11$ & $\boldsymbol{20.99 \pm 30.63}$ & 30/35 & $0.40 \pm 0.12$ \\
	        TinyLight & $-8.43 \pm 38.60$ & 21/35 & $0.64 \pm 0.16$ & $-25.19 \pm 34.56$ & 8/35 & $0.61 \pm 0.12$ \\
	        \bottomrule
	      \end{tabular}

	    \end{threeparttable}
	  \end{table*}

		  For RQ1, PPL-TSC achieves competitive closed-loop performance relative to its teachers for most teacher families. Tailored obtains a positive mean $\Delta$ATT for CoLight, FRAP, MPLight, and EcoLight (phase-selection), while Unified improves CoLight, FRAP, and EcoLight (phase-selection) and remains nearly neutral for MPLight. These results indicate that compact programmatic policies can preserve, and often improve, the teacher's control quality in the original deployment setting.

    The main performance exceptions occur when PPL-TSC learns from TinyLight and EcoLight (native). The student policy distilled from TinyLight underperforms its teacher on average, especially under Unified. TinyLight's richer phase-level and incoming-lane inputs enlarge the search space, making the default search budget less effective. For EcoLight (native), the two-action keep/switch interface makes closed-loop performance sensitive to accumulated switching errors, leading to weaker ATT results especially in larger networks such as Jinan, Los Angeles, and Manhattan (Appendix~\ref{app:scenario_level_results}). Controlled diagnostic experiments for these two cases are provided in Appendix~\ref{app:teacher_specific_diagnosis}.

          For RQ2, PPL-TSC maintains moderate behavioral fidelity to the teacher policies. Tailored consistently achieves higher ABF than Unified across teacher families, indicating that fitting a policy to each teacher--scenario--flow setting tends to better match local phase decisions than using a single policy across multiple scenarios. EcoLight (native) achieves the highest ABF under both Tailored ($0.84 \pm 0.07$) and Unified ($0.82 \pm 0.07$) settings, although its ATT results remain weaker than those of the teacher. Among the phase-selection teachers, MPLight provides the strongest fidelity--performance balance: both variants stay close to the teacher in ATT while achieving high ABF values, suggesting that its decision logic is more amenable to approximation by a compact programmatic policy. Detailed scenario-level results are provided in Appendix~\ref{app:scenario_level_results}.

  \subsection{Generalizability (RQ3)}
  We evaluate generalizability under two transfer regimes: (i) {same-scenario, different-traffic-flow}, where a model trained on one traffic flow is tested on other traffic flows within the same scenario, and (ii) {cross-scenario}, where a model learned in one scenario is applied to different scenarios. For same-scenario, different-traffic-flow transfer, each scenario contains five traffic flows; each traffic flow is used once as the source flow and evaluated on the other four flows, yielding 20 transfer cases per scenario and 140 cases per teacher. For cross-scenario transfer, each source traffic flow is evaluated on five traffic flows from each of the other six scenarios, yielding 150 transfer cases per source scenario and 1050 cases per teacher. Table~\ref{tab:generalization} reports relative ATT improvements and ATT win counts with respect to the transferred teacher policies.

  \begin{table*}[h]
    \centering
    \caption{Generalization under two transfer settings. Higher $\Delta$ATT and ATT wins are better. CoLight and TinyLight are not applicable to cross-scenario transfer due to architectural limitations.}
    \label{tab:generalization}
    \scriptsize
    \begin{threeparttable}
      \begin{minipage}[t]{0.47\textwidth}
        \centering
        \setlength{\tabcolsep}{5pt}
        \begin{tabular}{@{}lcc@{}}
          \toprule
          \multicolumn{3}{c}{\textbf{(a) Same-scenario, different-traffic-flow}} \\
          \midrule
          \textbf{Teacher} & $\Delta$ATT (\%) $\uparrow$ & ATT wins $\uparrow$ \\
          \midrule
	          CoLight & $\boldsymbol{15.66 \pm 38.49}$ & 61/140 \\
		          EcoLight (native) & $\boldsymbol{24.51 \pm 28.37}$ & 122/140 \\
		          EcoLight (phase-selection) & $\boldsymbol{25.19 \pm 31.00}$ & 120/140 \\
		          FRAP & $\boldsymbol{11.84 \pm 36.96}$ & 89/140 \\
		          MPLight & $\boldsymbol{0.57 \pm 11.79}$ & 55/140 \\
		          TinyLight & $-4.48 \pm 50.05$ & 98/140 \\
	          \bottomrule
	        \end{tabular}
      \end{minipage}\hspace{0.07\textwidth}
      \begin{minipage}[t]{0.49\textwidth}
        \centering
        \setlength{\tabcolsep}{5pt}
        \begin{tabular}{@{}lcc@{}}
          \toprule
          \multicolumn{3}{c}{\textbf{(b) Cross-scenario}} \\
          \midrule
          \textbf{Teacher} & $\Delta$ATT (\%) $\uparrow$ & ATT wins $\uparrow$ \\
          \midrule
			          EcoLight (native) & $\boldsymbol{33.42 \pm 77.99}$ & 938/1050 \\
			          EcoLight (phase-selection) & $\boldsymbol{44.85 \pm 34.32}$ & 954/1050 \\
		          FRAP & $\boldsymbol{54.22 \pm 39.35}$ & 887/1050 \\
		          MPLight & $\boldsymbol{46.04 \pm 36.10}$ & 946/1050 \\
	          \bottomrule
	        \end{tabular}
      \end{minipage}
    \end{threeparttable}
  \end{table*}

		  Table~\ref{tab:generalization} shows different patterns across the two transfer regimes. Under same-scenario, different-flow transfer, PPL-TSC achieves the strongest gains for CoLight, EcoLight (native), EcoLight (phase-selection), and FRAP, indicating that the learned policies generalize beyond the source flow under moderate flow shifts. Cross-scenario transfer is more challenging due to changes in both network topology and traffic patterns. In this setting, programmatic students for EcoLight (native), EcoLight (phase-selection), FRAP, and MPLight achieve substantial improvements over the transferred teachers. These comparisons measure gains over transferred teachers, whose performance can itself deteriorate under distribution shift; they do not establish absolute control quality on the target networks. A separate comparison using MPLight rollouts evaluates PPL-TSC against $\pi$-Light-Distill, GPLight+-Distill, and SymLight-Distill under the same teacher-guided protocol (Appendix~\ref{app:distill_baselines}). PPL-TSC achieves the lowest mean and median cross-scenario ATT among these methods, providing additional evidence of transfer effectiveness beyond the teacher comparison.
  
  


\subsection{Readability of the Learned Student Policy (RQ4)}
\label{sec:readability}

  \begin{figure*}[t]
    \centering
    \begin{minipage}[t]{0.55\textwidth}
      \vspace{0pt}
      \centering
      \includegraphics[width=\linewidth]{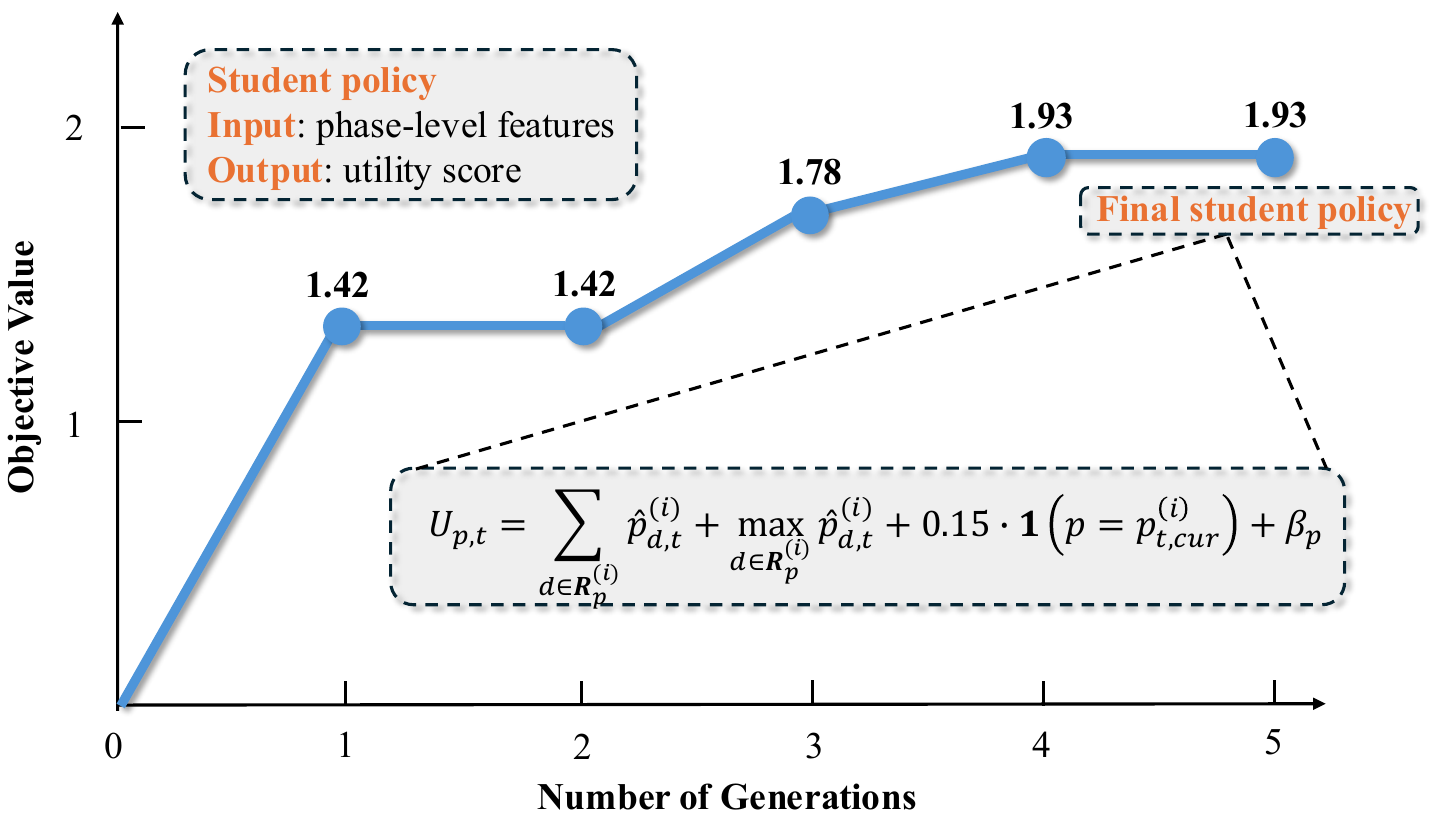}
      \captionof{figure}{Evolution of student (i.e., PPL-MPLight) on Hangzhou3. $\hat{p}_{d,t}$ denotes normalized pressure, $\mathcal{R}_{p}$ the set of served directions, $p_{t,\mathrm{cur}}$ the current phase, and $\beta_p$ the phase bias. The objective value is the joint evaluation score in Eq.~\eqref{eq:evaluation}.}
      \label{fig:student_policy_evo}
    \end{minipage}\hspace{0.07\textwidth}
    \begin{minipage}[t]{0.35\textwidth}
      \vspace{1.5em}
      \centering
      \captionof{table}{Final phase biases and control performance comparison on Hangzhou3 (single intersection with four phases).}
      \label{tab:student_policy_summary}
      \scriptsize
      \vspace{0.25em}
        \begin{tabular}{@{}p{2.5cm} c@{}}
        \toprule
        Quantity & Value \\
        \midrule
        Phase-0 bias $\beta_0$ & 0.42 \\
        Phase-1 bias $\beta_1$ & 0.42 \\
        Phase-2 bias $\beta_2$ & 0.24 \\
        Phase-3 bias $\beta_3$ & -0.34 \\
        \midrule
        Teacher ATT & 86.01 \\
        Student ATT & 84.04 \\
        MaxPressure ATT & 103.98 \\
        $\Delta$ATT (\%) & 2.29 \\
        ABF & 0.93 \\
        MaxPressure ABF & 0.81 \\
        \bottomrule
      \end{tabular}
    \end{minipage}
  \end{figure*}
  We examine the MPLight-derived student, PPL-MPLight, on Hangzhou3. The student uses the teacher's current-phase and movement-pressure inputs. Fig.~\ref{fig:student_policy_evo} traces the learned rule, and Table~\ref{tab:student_policy_summary} reports its calibrated biases and performance. The learned policy is readable because each term has a clear traffic-control interpretation. Utility increases with both aggregate phase pressure and the maximum directional pressure within a phase, giving priority to phases serving congested movements. A small current-phase retention term discourages unnecessary switching, and the calibrated biases provide phase-specific offsets around the shared pressure-based rule.

  This rule is closely related to MaxPressure \citep{varaiya2013max}, a classic pressure-based controller that selects phases using upstream--downstream queue differences. In this example, the student achieves both lower ATT (84.04 vs.\ 103.98) and higher behavioral fidelity to the teacher (ABF 0.93 vs.\ 0.81) than MaxPressure (Table~\ref{tab:student_policy_summary}). The explicit rule thus preserves recognizable pressure-based control logic while improving on MaxPressure in this case.

  \subsection{Resource Efficiency (RQ5)}
\begin{figure*}[t]
    \centering
    \begin{minipage}[t]{0.48\textwidth}
      \centering
      \includegraphics[width=\linewidth]{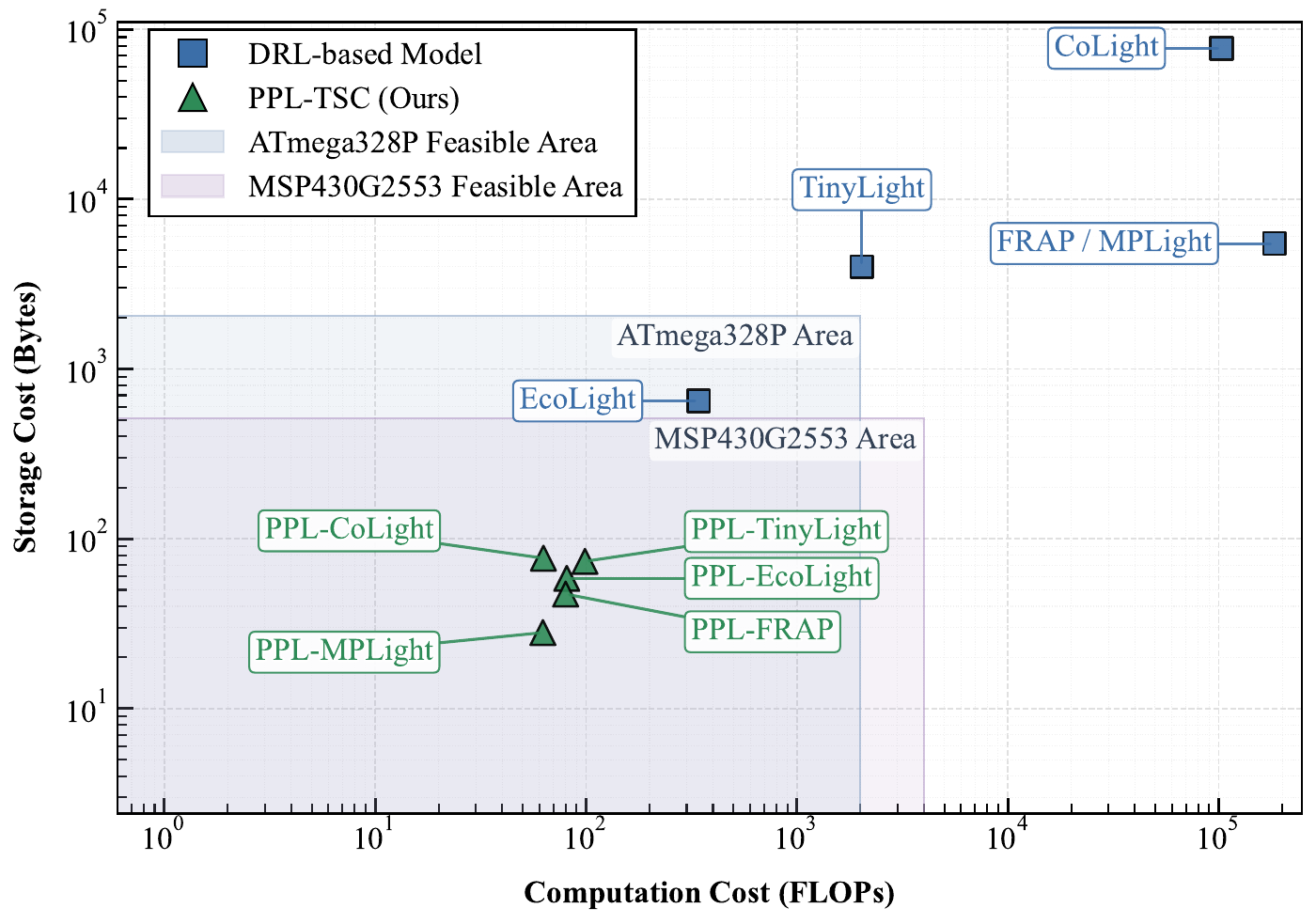}
      \captionof{figure}{Parameter storage and policy-scoring FLOPs on a representative Hangzhou3 intersection. Shaded microcontroller regions serve as resource-budget references, not verified deployment limits.}
      \label{fig:resource_efficiency_hangzhou3}
    \end{minipage}\hfill
    \begin{minipage}[t]{0.48\textwidth}
      \centering
      \includegraphics[width=\linewidth]{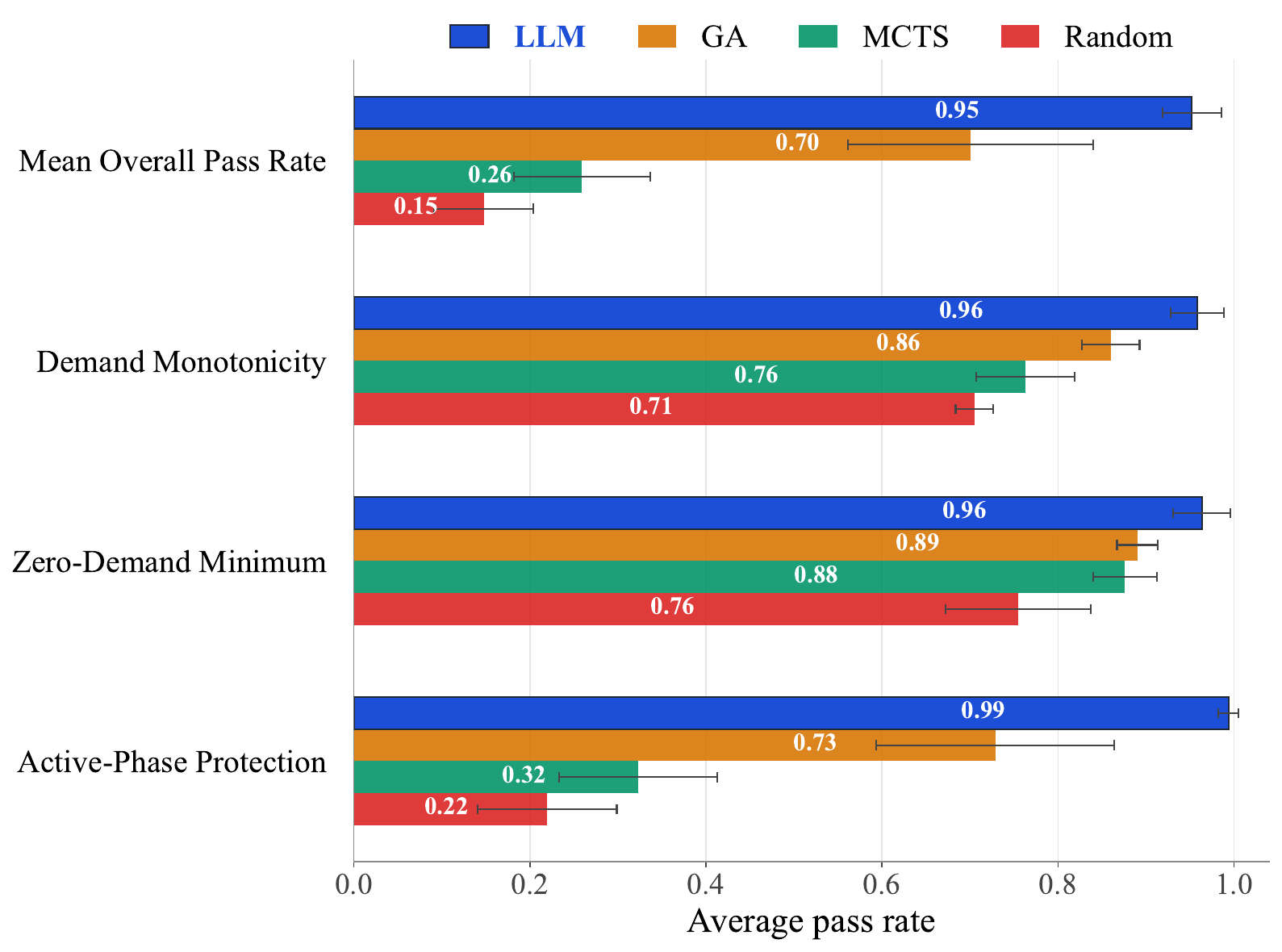}
      \captionof{figure}{Average pass rate under three semantic constraints for functions generated by LLM, GA, MCTS, and random search.}
      \label{fig:llm_mixed_source_ablation}
    \end{minipage}
  \end{figure*}

  We evaluate resource efficiency as an additional deployment benefit of the explicit programmatic representation, using two deployment-oriented metrics \citep{liao2025symlight,xing2022tinylight}: mean parameter storage and computational cost. Mean parameter storage is measured in bytes from the number of stored scalar parameters or rule constants, assuming 32-bit values. Computational cost is measured by the estimated floating-point operations (FLOPs) required for one intersection-level signal decision, counting the forward pass of neural teachers or the evaluation of phase utility functions over candidate phases. These metrics quantify parameter storage and policy-scoring operations; executable code, runtime memory, and feature preprocessing incur additional costs. Fig.~\ref{fig:resource_efficiency_hangzhou3} compares DRL teachers and the programmatic students learned by PPL-TSC on a representative Hangzhou3 intersection. The shaded regions for ATmega328P and MSP430G2553 provide illustrative resource-budget references; actual deployment also depends on implementation overhead and the required decision interval.

  As shown in Fig.~\ref{fig:resource_efficiency_hangzhou3}, the programmatic students consistently lie in the low-resource regime. Compared with their DRL teachers, they typically reduce both storage and FLOPs by one to three orders of magnitude. For example, FRAP-derived students reduce mean parameter storage from 5{,}476~bytes to 47.2~bytes and require an average of 80.0 FLOPs, compared with 183{,}676 for the teacher. For MPLight, the derived students require only 28.0~bytes and 62.4 FLOPs on average. TinyLight-derived students are also highly compact, achieving storage and FLOP reductions of 98.16\% and 95.14\%, respectively. Overall, these results establish lower storage and inference costs as an additional practical benefit of the programmatic representation, particularly for resource-constrained deployment.

  \subsection{Ablation Study (RQ6)}
  To answer RQ6, we perform targeted ablations on three components of PPL-TSC: policy generation, policy calibration, and policy evaluation. The calibration and evaluation ablations use the Tailored setting over 175 tasks: five teacher interfaces (CoLight, FRAP, MPLight, EcoLight (phase-selection), and TinyLight), seven scenarios, and five traffic flows.

  We first examine policy generation. Fig.~\ref{fig:llm_mixed_source_ablation} compares LLM, GA \citep{liao2025gplight+}, MCTS \citep{gu2024pi}, and random search under a matched time budget in terms of their ability to produce expressions satisfying three traffic-related semantic constraints: demand monotonicity, zero-demand minimum, and active-phase protection. Definitions of these constraints are provided in Appendix~\ref{app:constraints}. We report the mean overall pass rate, defined as the fraction of expressions satisfying all three constraints simultaneously. LLM achieves the highest observed pass rate on each constraint and on all three jointly.

  \begin{table*}[h]
    \centering
    \caption{Calibration and evaluation ablations in the Tailored setting over 175 tasks. The EcoLight results use the phase-selection interface, denoted EcoLight (phase-selection).}
    \label{tab:ablation}
    \scriptsize
    \begin{threeparttable}
      \begin{minipage}[t]{0.49\textwidth}
        \centering
        \setlength{\tabcolsep}{4pt}
        \begin{tabular}{@{}lccc@{}}
          \toprule
          \multicolumn{4}{c}{\textbf{(a) Policy calibration}} \\
          \midrule
          \textbf{Variant} & \textbf{$\Delta$ATT (\%) $\uparrow$} & \textbf{ATT wins $\uparrow$} & \textbf{ABF $\uparrow$} \\
          \midrule
          Full PPL-TSC & 7.62 $\pm$ 28.80 & 113/175 & 0.64$\pm$0.14  \\
          w/o calibration & 9.78 $\pm$ 34.96 & 116/175 & 0.55$\pm$0.18 \\
          \bottomrule
        \end{tabular}
      \end{minipage}\hfill
      \begin{minipage}[t]{0.47\textwidth}
        \centering
        \setlength{\tabcolsep}{4pt}
        \begin{tabular}{@{}lccc@{}}
          \toprule
          \multicolumn{4}{c}{\textbf{(b) Policy evaluation}} \\
          \midrule
          \textbf{Variant} & \textbf{$\Delta$ATT (\%) $\uparrow$} & \textbf{ATT wins $\uparrow$} & \textbf{ABF $\uparrow$} \\
          \midrule
          Full PPL-TSC & 7.62 $\pm$ 28.80 & 113/175 & 0.64$\pm$0.14\\
          Fidelity only & -14.07 $\pm$ 35.02 & 50/175 &  0.68$\pm$0.15 \\
          Performance only &12.82 $\pm$ 28.77  & 116/175 & 0.55$\pm$0.18 \\
          \bottomrule
        \end{tabular}
      \end{minipage}
    \end{threeparttable}
  \end{table*}
  
  We then analyze policy calibration (Table~\ref{tab:ablation}(a)). Removing calibration slightly improves ATT-oriented metrics, but reduces ABF from 0.64 to 0.55. This pattern suggests that, in our setting, calibration mainly improves alignment with the teacher's decisions rather than traffic efficiency directly. Table~\ref{tab:ablation}(b) compares variants of the policy objective in Eq.~\eqref{eq:evaluation}. The fidelity-only objective yields the highest ABF but substantially degrades ATT. The performance-only objective improves ATT metrics but reduces ABF. The joint objective provides a compromise between the two single-objective variants: it achieves higher ABF than performance-only evaluation and better ATT outcomes than fidelity-only evaluation.

\section{Conclusion}

We present PPL-TSC, a teacher-guided framework for learning human-readable programmatic student policies from DRL-based traffic signal controllers. Across diverse scenarios and teacher families, PPL-TSC achieves competitive closed-loop performance while maintaining moderate behavioral fidelity to its teachers. The learned policies show strong transfer performance under traffic-flow shifts and cross-scenario deployment, and substantially reduce storage and computational costs relative to neural teachers. Future work will extend PPL-TSC to broader TSC settings and evaluate its robustness in more realistic large-scale deployments.

\begingroup
\renewcommand{\acksection}{\section*{Acknowledgments}}
\begin{ack}
This work was supported by the Mainland-Hong Kong Joint Funding Scheme of the Innovation and Technology Commission (Grant No. MHP/038/23) and the General Research Fund of the Research Grants Council of the Hong Kong Special Administrative Region, China (Grant No. 15224423). The authors declare no competing interests.
\end{ack}
\endgroup

\bibliographystyle{unsrtnat} 
\bibliography{References}

\begin{thebibliography}{36}
\providecommand{\natexlab}[1]{#1}
\providecommand{\url}[1]{\texttt{#1}}
\expandafter\ifx\csname urlstyle\endcsname\relax
  \providecommand{\doi}[1]{doi: #1}\else
  \providecommand{\doi}{doi: \begingroup \urlstyle{rm}\Url}\fi

\bibitem[Wei et~al.(2019{\natexlab{a}})Wei, Zheng, Gayah, and
  Li]{wei2019survey}
Hua Wei, Guanjie Zheng, Vikash Gayah, and Zhenhui Li.
\newblock A survey on traffic signal control methods.
\newblock \emph{arXiv preprint arXiv:1904.08117}, 2019{\natexlab{a}}.

\bibitem[Wu et~al.(2025)Wu, Ding, Lin, Zheng, Sun, Fang, Xu, Zhu, and
  Gu]{wu2025big}
Kan Wu, Jianrong Ding, Jingli Lin, Guanjie Zheng, Yi~Sun, Jie Fang, Tu~Xu,
  Yongdong Zhu, and Baojing Gu.
\newblock Big-data empowered traffic signal control could reduce urban carbon
  emission.
\newblock \emph{Nature Communications}, 16\penalty0 (1):\penalty0 2013, 2025.

\bibitem[Webster(1958)]{webster1958traffic}
Fo~Vo Webster.
\newblock Traffic signal settings.
\newblock Technical report, 1958.

\bibitem[Little(1966)]{little1966synchronization}
John~DC Little.
\newblock The synchronization of traffic signals by mixed-integer linear
  programming.
\newblock \emph{Operations Research}, 14\penalty0 (4):\penalty0 568--594, 1966.

\bibitem[Little et~al.(1981)Little, Kelson, and Gartner]{little1981maxband}
John~DC Little, Mark~D Kelson, and Nathan~H Gartner.
\newblock {MAXBAND}: A versatile program for setting signals on arteries and
  triangular networks.
\newblock 1981.

\bibitem[Cools et~al.(2007)Cools, Gershenson, and D’Hooghe]{cools2007self}
Seung-Bae Cools, Carlos Gershenson, and Bart D’Hooghe.
\newblock Self-organizing traffic lights: A realistic simulation.
\newblock In \emph{Advances in Applied Self-Organizing Systems}, pages 41--50.
  Springer, 2007.

\bibitem[Varaiya(2013)]{varaiya2013max}
Pravin Varaiya.
\newblock The max-pressure controller for arbitrary networks of signalized
  intersections.
\newblock In \emph{Advances in Dynamic Network Modeling in Complex
  Transportation Systems}, pages 27--66. Springer, 2013.

\bibitem[Hunt et~al.(1981)Hunt, Robertson, Bretherton, and
  Winton]{hunt1981scoot}
PB~Hunt, DI~Robertson, RD~Bretherton, and RI~Winton.
\newblock {SCOOT}-a traffic responsive method of coordinating signals.
\newblock Technical report, 1981.

\bibitem[Lowrie(1990)]{lowrie1990scats}
PR~Lowrie.
\newblock {SCATS}, {Sydney} co-ordinated adaptive traffic system: A traffic
  responsive method of controlling urban traffic.
\newblock 1990.

\bibitem[Wei et~al.(2021)Wei, Zheng, Gayah, and Li]{wei2021recent}
Hua Wei, Guanjie Zheng, Vikash Gayah, and Zhenhui Li.
\newblock Recent advances in reinforcement learning for traffic signal control:
  A survey of models and evaluation.
\newblock \emph{ACM SIGKDD Explorations Newsletter}, 22\penalty0 (2):\penalty0
  12--18, 2021.

\bibitem[Li et~al.(2016)Li, Lv, and Wang]{li2016traffic}
Li~Li, Yisheng Lv, and Fei-Yue Wang.
\newblock Traffic signal timing via deep reinforcement learning.
\newblock \emph{IEEE/CAA Journal of Automatica Sinica}, 3\penalty0
  (3):\penalty0 247--254, 2016.

\bibitem[Yau et~al.(2017)Yau, Qadir, Khoo, Ling, and
  Komisarczuk]{yau2017survey}
Kok-Lim~Alvin Yau, Junaid Qadir, Hooi~Ling Khoo, Mee~Hong Ling, and Peter
  Komisarczuk.
\newblock A survey on reinforcement learning models and algorithms for traffic
  signal control.
\newblock \emph{ACM Computing Surveys (CSUR)}, 50\penalty0 (3):\penalty0 1--38,
  2017.

\bibitem[Xiao et~al.(2025)Xiao, Lu, Li, Tu, and Li]{xiao2025advances}
Feng Xiao, Jiaming Lu, Lu~Li, Wenwen Tu, and Chaojing Li.
\newblock Advances in reinforcement learning for traffic signal control: a
  review of recent progress.
\newblock \emph{Intelligent Transportation Infrastructure}, 4:\penalty0
  liaf009, 2025.

\bibitem[Wei et~al.(2019{\natexlab{b}})Wei, Xu, Zhang, Zheng, Zang, Chen,
  Zhang, Zhu, Xu, and Li]{wei2019colight}
Hua Wei, Nan Xu, Huichu Zhang, Guanjie Zheng, Xinshi Zang, Chacha Chen, Weinan
  Zhang, Yanmin Zhu, Kai Xu, and Zhenhui Li.
\newblock {CoLight}: Learning network-level cooperation for traffic signal
  control.
\newblock In \emph{Proceedings of the 28th ACM International Conference on
  Information and Knowledge Management}, pages 1913--1922, 2019{\natexlab{b}}.

\bibitem[Zheng et~al.(2019)Zheng, Xiong, Zang, Feng, Wei, Zhang, Li, Xu, and
  Li]{zheng2019learning}
Guanjie Zheng, Yuanhao Xiong, Xinshi Zang, Jie Feng, Hua Wei, Huichu Zhang,
  Yong Li, Kai Xu, and Zhenhui Li.
\newblock Learning phase competition for traffic signal control.
\newblock In \emph{Proceedings of the 28th ACM International Conference on
  Information and Knowledge Management}, pages 1963--1972, 2019.

\bibitem[Zhu et~al.(2022)Zhu, Yin, and Chen]{zhu2022extracting}
Yuanyang Zhu, Xiao Yin, and Chunlin Chen.
\newblock Extracting decision tree from trained deep reinforcement learning in
  traffic signal control.
\newblock \emph{IEEE Transactions on Computational Social Systems}, 10\penalty0
  (4):\penalty0 1997--2007, 2022.

\bibitem[Gu et~al.(2024)Gu, Zhang, Liu, Gao, Li, and Zhou]{gu2024pi}
Yin Gu, Kai Zhang, Qi~Liu, Weibo Gao, Longfei Li, and Jun Zhou.
\newblock {$\pi$-Light}: Programmatic interpretable reinforcement learning for
  resource-limited traffic signal control.
\newblock In \emph{Proceedings of the AAAI Conference on Artificial
  Intelligence}, volume~38, pages 21107--21115, 2024.

\bibitem[Chauhan et~al.(2020)Chauhan, Bansal, and Sen]{chauhan2020ecolight}
Sachin Chauhan, Kashish Bansal, and Rijurekha Sen.
\newblock {EcoLight}: Intersection control in developing regions under extreme
  budget and network constraints.
\newblock \emph{Advances in Neural Information Processing Systems},
  33:\penalty0 13027--13037, 2020.

\bibitem[Chen et~al.(2020)Chen, Wei, Xu, Zheng, Yang, Xiong, Xu, and
  Li]{chen2020toward}
Chacha Chen, Hua Wei, Nan Xu, Guanjie Zheng, Ming Yang, Yuanhao Xiong, Kai Xu,
  and Zhenhui Li.
\newblock Toward a thousand lights: Decentralized deep reinforcement learning
  for large-scale traffic signal control.
\newblock In \emph{Proceedings of the AAAI Conference on Artificial
  Intelligence}, volume~34, pages 3414--3421, 2020.

\bibitem[Xing et~al.(2022)Xing, Zheng, Liu, and Pan]{xing2022tinylight}
Dong Xing, Qian Zheng, Qianhui Liu, and Gang Pan.
\newblock {TinyLight}: Adaptive traffic signal control on devices with
  extremely limited resources.
\newblock \emph{arXiv preprint arXiv:2205.00427}, 2022.

\bibitem[Liao et~al.(2025{\natexlab{a}})Liao, Mei, and Zhang]{liao2025gplight+}
Xiao-Cheng Liao, Yi~Mei, and Mengjie Zhang.
\newblock {GPLight+}: a genetic programming method for learning symmetric
  traffic signal control policy.
\newblock \emph{IEEE Transactions on Evolutionary Computation},
  2025{\natexlab{a}}.

\bibitem[Liao et~al.(2025{\natexlab{b}})Liao, Mei, and Zhang]{liao2025symlight}
Xiao-Cheng Liao, Yi~Mei, and Mengjie Zhang.
\newblock {SymLight}: Exploring interpretable and deployable symbolic policies
  for traffic signal control.
\newblock \emph{arXiv preprint arXiv:2511.05790}, 2025{\natexlab{b}}.

\bibitem[{\'S}wiechowski et~al.(2023){\'S}wiechowski, Godlewski, Sawicki, and
  Ma{\'n}dziuk]{swiechowski2023monte}
Maciej {\'S}wiechowski, Konrad Godlewski, Bartosz Sawicki, and Jacek
  Ma{\'n}dziuk.
\newblock {Monte Carlo} tree search: A review of recent modifications and
  applications.
\newblock \emph{Artificial Intelligence Review}, 56\penalty0 (3):\penalty0
  2497--2562, 2023.

\bibitem[Koza(1990)]{koza1990genetic}
John~R Koza.
\newblock \emph{Genetic programming: A paradigm for genetically breeding
  populations of computer programs to solve problems}, volume~34.
\newblock Stanford University, Department of Computer Science Stanford, CA,
  1990.

\bibitem[Aslani et~al.(2017)Aslani, Mesgari, and Wiering]{aslani2017adaptive}
Mohammad Aslani, Mohammad~Saadi Mesgari, and Marco Wiering.
\newblock Adaptive traffic signal control with actor-critic methods in a
  real-world traffic network with different traffic disruption events.
\newblock \emph{Transportation Research Part C: Emerging Technologies},
  85:\penalty0 732--752, 2017.

\bibitem[Rizzo et~al.(2019)Rizzo, Vantini, and Chawla]{rizzo2019time}
Stefano~Giovanni Rizzo, Giovanna Vantini, and Sanjay Chawla.
\newblock Time critic policy gradient methods for traffic signal control in
  complex and congested scenarios.
\newblock In \emph{Proceedings of the 25th ACM SIGKDD International Conference
  on Knowledge Discovery \& Data Mining}, pages 1654--1664, 2019.

\bibitem[Chu et~al.(2019)Chu, Wang, Codec{\`a}, and Li]{chu2019multi}
Tianshu Chu, Jie Wang, Lara Codec{\`a}, and Zhaojian Li.
\newblock Multi-agent deep reinforcement learning for large-scale traffic
  signal control.
\newblock \emph{IEEE Transactions on Intelligent Transportation Systems},
  21\penalty0 (3):\penalty0 1086--1095, 2019.

\bibitem[Oroojlooy et~al.(2020)Oroojlooy, Nazari, Hajinezhad, and
  Silva]{oroojlooy2020attendlight}
Afshin Oroojlooy, Mohammadreza Nazari, Davood Hajinezhad, and Jorge Silva.
\newblock {AttendLight}: Universal attention-based reinforcement learning model
  for traffic signal control.
\newblock \emph{Advances in Neural Information Processing Systems},
  33:\penalty0 4079--4090, 2020.

\bibitem[Ault et~al.(2019)Ault, Hanna, and Sharon]{ault2019learning}
James Ault, Josiah~P Hanna, and Guni Sharon.
\newblock Learning an interpretable traffic signal control policy.
\newblock \emph{arXiv preprint arXiv:1912.11023}, 2019.

\bibitem[Ahvanooey et~al.(2019)Ahvanooey, Li, Wu, and
  Wang]{ahvanooey2019survey}
Milad~Taleby Ahvanooey, Qianmu Li, Ming Wu, and Shuo Wang.
\newblock A survey of genetic programming and its applications.
\newblock \emph{KSII Trans. Internet Inf. Syst.}, 13\penalty0 (4):\penalty0
  1765--1794, 2019.

\bibitem[Xie et~al.(2025)Xie, Zhang, Tang, Chung, Li, and Wang]{xie2025co}
Lindong Xie, Yang Zhang, Zhixian Tang, Edward Chung, Genghui Li, and Zhenkun
  Wang.
\newblock Co-evolution of large language models and configuration strategies to
  enhance surrogate-assisted evolutionary algorithm.
\newblock In \emph{Proceedings of the 31st ACM SIGKDD Conference on Knowledge
  Discovery and Data Mining V. 2}, pages 3321--3332, 2025.

\bibitem[Liu et~al.(2024)Liu, Tong, Yuan, Lin, Luo, Wang, Lu, and
  Zhang]{liu2024evolution}
Fei Liu, Xialiang Tong, Mingxuan Yuan, Xi~Lin, Fu~Luo, Zhenkun Wang, Zhichao
  Lu, and Qingfu Zhang.
\newblock Evolution of heuristics: Towards efficient automatic algorithm design
  using large language model.
\newblock \emph{arXiv preprint arXiv:2401.02051}, 2024.

\bibitem[Joel et~al.(2024)Joel, Wu, and Fard]{joel2024survey}
Sathvik Joel, Jie Wu, and Fatemeh Fard.
\newblock A survey on {LLM}-based code generation for low-resource and
  domain-specific programming languages.
\newblock \emph{ACM Transactions on Software Engineering and Methodology},
  2024.

\bibitem[Hansen et~al.(2003)Hansen, M{\"u}ller, and
  Koumoutsakos]{hansen2003reducing}
Nikolaus Hansen, Sibylle~D M{\"u}ller, and Petros Koumoutsakos.
\newblock Reducing the time complexity of the derandomized evolution strategy
  with covariance matrix adaptation ({CMA-ES}).
\newblock \emph{Evolutionary Computation}, 11\penalty0 (1):\penalty0 1--18,
  2003.

\bibitem[Zhang et~al.(2019)Zhang, Feng, Liu, Ding, Zhu, Zhou, Zhang, Yu, Jin,
  and Li]{zhang2019cityflow}
Huichu Zhang, Siyuan Feng, Chang Liu, Yaoyao Ding, Yichen Zhu, Zihan Zhou,
  Weinan Zhang, Yong Yu, Haiming Jin, and Zhenhui Li.
\newblock {CityFlow}: A multi-agent reinforcement learning environment for
  large scale city traffic scenario.
\newblock In \emph{The World Wide Web Conference}, pages 3620--3624.
  Association for Computing Machinery, 2019.
\newblock \doi{10.1145/3308558.3314139}.

\bibitem[Mei et~al.(2024)Mei, Lei, Da, Shi, and Wei]{mei2024libsignal}
Hao Mei, Xiaoliang Lei, Longchao Da, Bin Shi, and Hua Wei.
\newblock {LibSignal}: an open library for traffic signal control.
\newblock \emph{Machine Learning}, 113\penalty0 (8):\penalty0 5235--5271, 2024.

\end{thebibliography}

\appendix
\section{Appendix}
\renewcommand{\thefigure}{A.\arabic{figure}}
\renewcommand{\thetable}{A.\arabic{table}}
\setcounter{figure}{0}
\setcounter{table}{0}

\subsection{Preliminaries}
\subsubsection{Term Definitions}
\label{app:definitions}

\begin{figure}[htbp]
\centering
\includegraphics[width=0.5\linewidth]{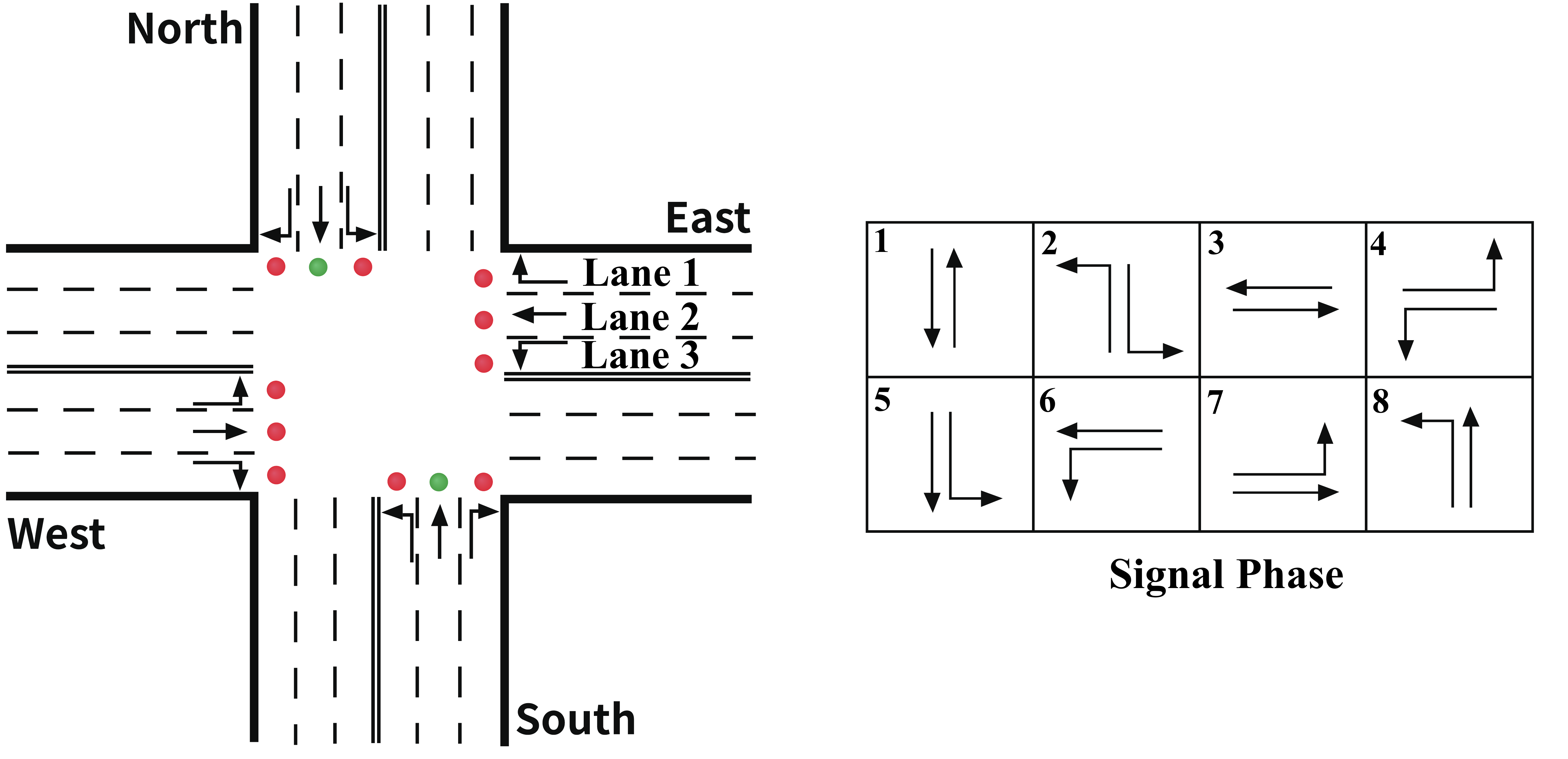}
\caption{Illustration of a four-approach signalized intersection with eight signal phases. Phase~1 is shown as the active phase, permitting one north-to-south through movement and one south-to-north through movement.}

\label{fig:Intersection_layout}
\end{figure}

We use a \textit{single signalized intersection} as an illustrative example to introduce several basic concepts in traffic signal control.

\textbf{1) Intersection, Approaches, and Lanes.}
As shown in Fig.~\ref{fig:Intersection_layout}, the intersection consists of four \textit{approaches} corresponding to the cardinal directions: north, east, south, and west. Each approach contains multiple \textit{incoming lanes}, where vehicles queue before crossing the stop line. In practice, lanes are assigned to different turning purposes, including \textit{left-turn}, \textit{through}, and \textit{right-turn} lanes. Vehicles are discharged from these lanes when the active signal phase permits the corresponding movements.

\textbf{2) Traffic Movements.}
A \textit{traffic movement} represents the trajectory of vehicles from an incoming lane to an outgoing lane. As illustrated by the arrows in Fig.~\ref{fig:Intersection_layout}, typical movements include \textit{left-turn}, \textit{through}, and \textit{right-turn} maneuvers. Different movements may conflict if their trajectories intersect within the intersection area, and must therefore be coordinated to ensure safety and efficiency.

\textbf{3) Phases.}
A \textit{phase} is a predefined set of \textit{non-conflicting traffic movements} that are allowed to proceed simultaneously. In the example in Fig.~\ref{fig:Intersection_layout}, eight signal phases are defined to coordinate different movements. During a given phase, vehicles associated with permitted movements receive a green signal, while conflicting movements remain stopped.

\textbf{4) Phase Switching.}
\textit{Phase switching} refers to the transition from one signal phase to another over time. For example, Fig.~\ref{fig:Intersection_layout} shows Phase~1 as the active phase, permitting the north-to-south and south-to-north through movements. At decision step $t$, intersection $i$ observes a local traffic state $\mathbf{s}_t^{(i)}$ and chooses a phase $p_t^{(i)}$ from a feasible candidate set $\mathcal{P}_t^{(i)}$, which may contain all phases or only the current phase and its designated successor. The objective is to choose which phase to activate, or whether to keep or switch the current phase, to improve traffic efficiency.

\textbf{5) Phase-Level Features.}
Each candidate phase $p$ is associated with a set of \emph{phase-level features} extracted from the current traffic state. Let $\mathcal{L}_p$ denote the set of incoming lanes associated with phase $p$, and let $\mathcal{M}_p$ denote the set of movements permitted by phase $p$. For each lane $\ell \in \mathcal{L}_p$ or movement $m \in \mathcal{M}_p$, base measurements such as queue length, waiting time, and vehicle count are available. We construct phase-level features by first mapping lane- or lane-link-level measurements to movement directions, such as north-to-south or east-to-west through movements, and then aggregating them over the movements served by phase $p$. This yields phase-level quantities such as pressure, vehicle counts, and waiting-related metrics. We further compute statistical descriptors, including $\{\mathrm{min}, \mathrm{max}, \mathrm{mean}, \mathrm{std}\}$, to capture distributional characteristics for normalization.

All aggregated and statistical features, together with the phase identifier, are concatenated to form the phase-level feature vector $\mathbf{x}_{p,t}^{(i)}$ for candidate phase $p$ at intersection $i$ and decision step $t$. This vector serves as the input to the utility function $f(\cdot)$.



\subsubsection{DRL-based Traffic Signal Control Methods}
\label{app:drl_background}
Deep reinforcement learning (DRL) has been widely adopted for traffic signal control (TSC) due to its ability to learn adaptive control policies directly from traffic data. In DRL-based TSC, the signal control problem is typically formulated as a Markov Decision Process (MDP), defined by the tuple $(\mathcal{S}, \mathcal{A}, \mathcal{K}, r, \gamma)$, where $r$ denotes the reward function.

The \textit{state} $s_t \in \mathcal{S}$ represents the current traffic condition at time step $t$, which may include features such as queue lengths, vehicle counts, or waiting times on incoming lanes. The \textit{action} $a_t \in \mathcal{A}$ corresponds to the control decision taken by the agent, typically selecting the next signal phase or determining whether to switch the current phase. 

After executing an action, the traffic environment transitions to a new state $s_{t+1}$ according to the transition kernel $\mathcal{K}(s_{t+1}\mid s_t,a_t)$. The agent then receives a scalar \textit{reward} $r_t$, obtained from $r$, which reflects the quality of the control decision, such as average travel time.

The objective of DRL-based traffic signal control is to learn a policy $\pi(a\mid s)$ that maximizes the expected discounted return over time:
\[
\max_{\pi} \ \mathbb{E}\left[\sum_{t=0}^{T} \gamma^t r_t \right],
\]
where $\gamma \in (0,1)$ is the discount factor and $T$ is the decision horizon. The learned policy enables the traffic signal controller to adaptively select signal phases based on real-time traffic conditions.


\subsection{Prompt Templates}
\label{app:prompt_templates}

\begin{figure}[htbp]
\centering
\includegraphics[width=1\linewidth]{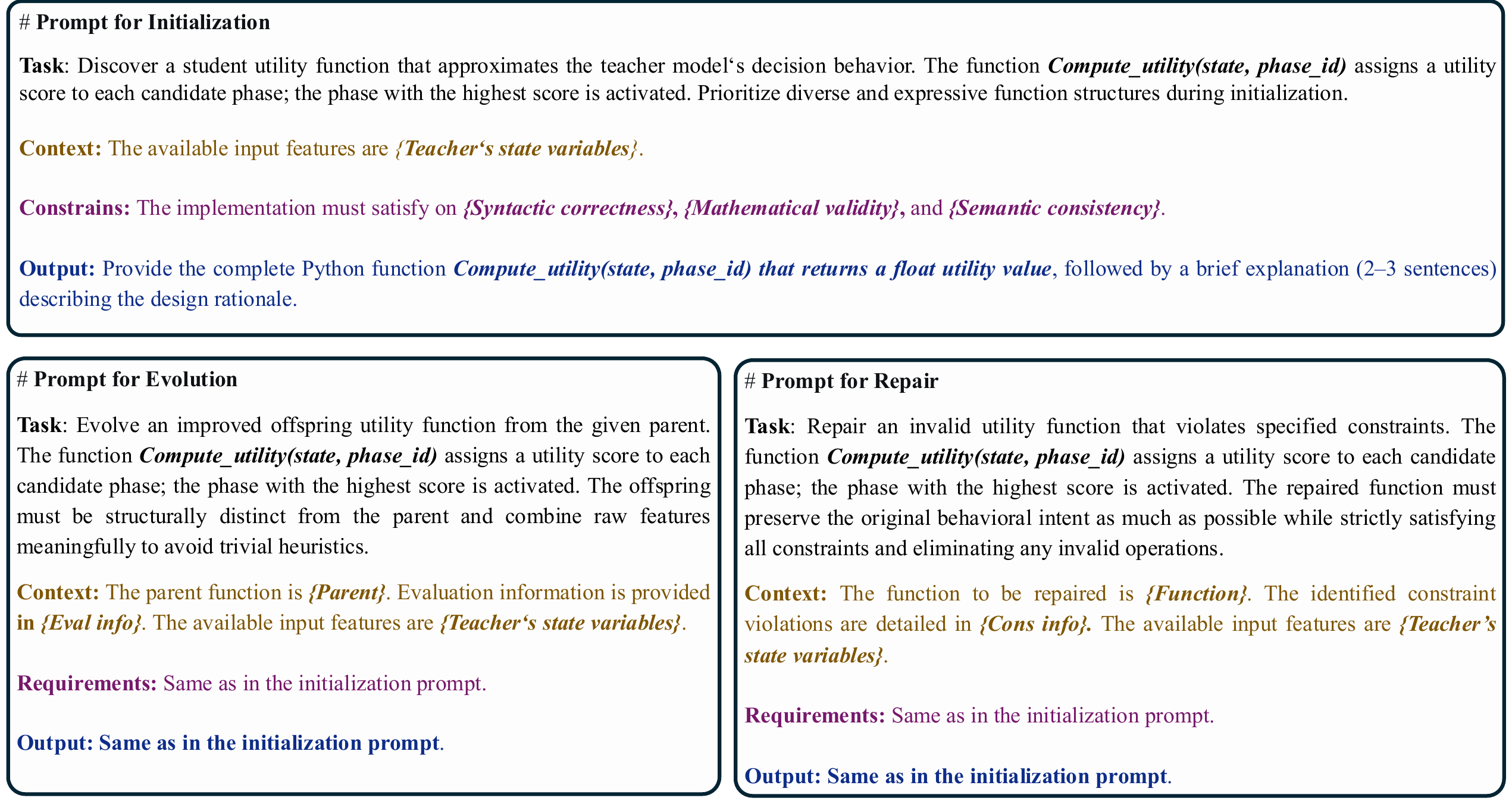}
\caption{Illustration of the initialization, evolution, and repair prompt templates used for LLM-based policy generation and revision.}
\label{fig:prompts}
\end{figure}

This subsection provides representative LLM prompt templates for policy generation and repair in PPL-TSC. Each template follows the structure $\mathcal{T}(\text{Task}, \text{Ctx}, \text{Cons}, \text{Out})$: task specifies the generation objective, context provides traffic features or parent PUFs, constraints encode the syntactic, mathematical, and semantic requirements used to guide generation, and output asks for an executable candidate PUF with a brief rationale. Fig.~\ref{fig:prompts} illustrates the initialization prompt, evolution prompt, and repair prompt.

\subsection{Details of Syntactic, Mathematical, and Semantic Constraints}
\label{app:constraints}

During policy review, each PUF produced by policy generation is explicitly checked against the same constraints encoded in the generation prompt before calibration. To ensure that generated utility functions are executable, numerically stable, and consistent with traffic-control principles, we enforce three families of validation checks: syntactic, mathematical, and semantic. An overview is provided in Table~\ref{tab:constraint_checks}.

\begin{table}[htbp]
\centering
\caption{Constraint validation checks.}
\label{tab:constraint_checks}
\scriptsize
\begin{tabularx}{\textwidth}{l l X}
\toprule
\textbf{Family} & \textbf{Check} & \textbf{Description} \\
\midrule
\multirow{3}{*}{Syntactic}
& Parseability & The generated function must have valid Python syntax and compile successfully. \\
& Executability & The generated function must be loadable and callable at runtime. \\
& Structural validity & The generated function must have a valid structure and return a value on every execution path. \\
\midrule
\multirow{3}{*}{Mathematical}
& Finite output & Output must be a finite real number (no \texttt{NaN} or infinite values). \\
& Domain safety & No invalid operations (e.g., division by zero, invalid logarithms, or square roots of negative inputs). \\
& Runtime robustness & No exceptions under real teacher states and extreme input scenarios (e.g., all-zero, all-max, mixed, single-feature extremes). \\
\midrule
\multirow{4}{*}{Semantic}
 & Feature legality & Only allowed traffic features are accessed. \\
 & Demand monotonicity & Phase utility or switching preference varies monotonically with demand according to the action interface. \\
 & Zero-demand minimum & With other inputs fixed, setting non-negative demand features to zero must not increase phase utility. \\
 & Active-phase protection & With demand fixed, the active-phase indicator must not reduce phase utility. \\
\bottomrule
\end{tabularx}
\end{table}
\paragraph{Syntactic checks.}
These checks ensure that the generated code is executable in the target pipeline. Each candidate must first pass Python compilation, including a valid function body and valid syntax. It is then loaded into the runtime environment via \texttt{exec()}, where the required entry point \texttt{compute\_utility} must exist and be callable. Malformed structures, such as invalid function skeletons or missing return statements, are rejected before acceptance.

\paragraph{Mathematical checks.}
These checks ensure numerical validity and stability. All outputs must be finite; any \texttt{NaN}, \texttt{inf}, or \texttt{-inf} results in rejection. All operations must respect valid input domains, such as non-zero denominators, positive inputs to logarithms, and non-negative inputs to square roots. The function must also avoid extreme intermediate values that could cause overflow or numerical instability. To test robustness, we evaluate each candidate on phase-level inputs constructed under the following scenarios:
\begin{itemize}
    \item \textbf{Real teacher states}: all samples $(\mathbf{s}_t^{(i)}, p_t^{\mathrm{Tch},(i)}) \in \mathcal{D}$, to ensure stability under the observed data distribution;
    \item \textbf{All-zero}: all traffic-measurement features set to zero, with phase indicators kept fixed;
    \item \textbf{All-max}: all traffic-measurement features set to their maximum observed values in $\mathcal{D}$, with phase indicators kept fixed;
    \item \textbf{Mixed-half}: half of the traffic-measurement features set to zero and the other half set to their maximum values;
    \item \textbf{Single-feature-max}: each traffic-measurement feature individually set to its maximum observed value, with all remaining features kept at their original sampled values.
\end{itemize}

\paragraph{Semantic checks.}
The feature whitelist covers the base traffic measurements \texttt{pressure}, \texttt{num\_vehicle}, \texttt{num\_waiting\_vehicle}, \texttt{waiting\_time}, and \texttt{sum\_waiting\_time}, together with \texttt{phase\_id} and \texttt{current\_phase}. Candidates may use the defined aggregation and normalization operations on these measurements to construct phase-level features. Access to undefined or disallowed inputs results in rejection.

\textbf{Demand monotonicity} follows the controller's action interface. For phase-local scoring, utility must be non-decreasing in the demand served by the candidate phase, with other inputs fixed. This applies to phase-selection policies and the original keep/switch student that scores the current and next phases separately. We evaluate this property on sampled real teacher states using multiple scaling factors. For each demand-related feature
\[
\begin{aligned}
x \in \{&\texttt{pressure},\ \texttt{num\_vehicle},\ \texttt{num\_waiting\_vehicle},\\
&\texttt{waiting\_time},\ \texttt{sum\_waiting\_time}\},
\end{aligned}
\]
we generate perturbed states with scaling factors $\alpha \in \{0.5, 0.8, 1.0, 1.2, 1.5\}$:
\[
x_\alpha = \alpha \cdot x_{\text{base}},
\]
while keeping all other features unchanged.

For a candidate utility function $f$, we evaluate the corresponding phase-level feature vector $\mathbf{x}_{p,t,\alpha}^{(i)}$ for each $\alpha$. We sort the perturbed feature values in ascending numerical order and check that utility is non-decreasing in that order:
\[
f(\mathbf{x}_{p,t,\alpha_k}^{(i)}) \geq f(\mathbf{x}_{p,t,\alpha_j}^{(i)}), \quad \forall x_{\alpha_k} > x_{\alpha_j}.
\]
Ordering by feature value rather than scaling factor accommodates signed features such as movement pressure. This check assesses monotonicity at the tested inputs.

For the aligned keep/switch variant in Appendix~\ref{app:teacher_specific_diagnosis}, we instead check the relative preference for switching. Let $g$ and $r$ denote green- and red-side demand, and define the switching margin as
\[
m(g,r)=f_{\mathrm{switch}}(g,r)-f_{\mathrm{keep}}(g,r).
\]
On sampled teacher states, we vary each demand input using the scaling factors above and hold the other inputs fixed. The margin must be non-increasing in $g$ and non-decreasing in $r$: greater green-side demand favors keeping the current phase, and greater red-side demand favors switching. These checks assess monotonicity at the tested inputs.

\textbf{Zero-demand minimum.} With other inputs fixed, setting non-negative demand features to zero must not increase phase utility. For aligned keep/switch policies, we use the switching-margin checks defined above.

\textbf{Active-phase protection} compares otherwise identical candidate inputs that differ only in whether the phase is currently active. The check verifies that the active-phase indicator is not used to penalize the current phase under comparable demand conditions. This discourages excessive phase switching and reflects the practical need for short-term signal stability.

\paragraph{Decision rule.}
Candidates that fail any check are returned to the LLM with diagnostic feedback and reviewed again after repair. Only candidates that pass all checks proceed to calibration and evaluation.

\subsection{Method Complexity Analysis}
\label{app:complexity}

We analyze the complexity of PPL-TSC in terms of both search-time cost and deployment-time inference cost. Let $|\mathcal{D}|=\sum_{i=1}^{M}N_i$ denote the total number of rollout samples, $B_i=\max_t|\mathcal{P}_t^{(i)}|$ the maximum feasible phase-set size at intersection $i$ (all phases for phase-selection controllers and $2$ for keep/switch controllers), $C_f$ the cost of evaluating $f(\cdot)$ on one phase-level feature vector, and $C_{\mathrm{LLM}}$ the cost of one LLM prompt-response interaction. We further use $G_{\max}$ for the maximum number of generations, $N_{\mathrm{pop}}$ for the population size, $R$ for the number of review inputs per candidate, $T_i$ for the number of CMA-ES objective evaluations at intersection $i$, $\bar{B}$ for the average candidate phase-set size, $T_{\mathrm{sim}}$ for the number of simulator control steps, $C_{\mathrm{env}}$ for one simulator transition, and $K_{\mathrm{rep}}$ for the average number of LLM generations per accepted candidate, including repair calls.

\begin{table}[htbp]
\centering
\caption{Complexity of the main steps in PPL-TSC.}
\label{tab:complexity_summary}
\scriptsize
\begin{tabularx}{0.92\textwidth}{l X}
\toprule
\textbf{Component} & \textbf{Complexity} \\
\midrule
Policy generation & $O(G_{\max}N_{\mathrm{pop}} K_{\mathrm{rep}} C_{\mathrm{LLM}})$ \\
Policy review & $O(R C_f)$ per reviewed candidate \\
Policy calibration & $O\!\left(\sum_{i=1}^{M} T_i N_i B_i C_f\right)$ per candidate \\
Policy evaluation & $O(1)$ for cached agreement; $O\!\left(T_{\mathrm{sim}}(M\bar{B}C_f + C_{\mathrm{env}})\right)$ for simulation \\
Deployment & $O(B_i C_f)$ per decision step \\
\bottomrule
\end{tabularx}
\end{table}

Table~\ref{tab:complexity_summary} summarizes the costs. Across the search, generation requires $O(G_{\max}N_{\mathrm{pop}}K_{\mathrm{rep}}C_{\mathrm{LLM}})$, and review requires $O(G_{\max}N_{\mathrm{pop}}K_{\mathrm{rep}}RC_f)$, excluding lightweight parsing overhead. LLM latency depends on prompt and response lengths and repair frequency.

\paragraph{Calibration and evaluation.}
One CMA-ES objective evaluation scores up to $B_i$ phases for each of $N_i$ samples, costing $O(N_iB_iC_f)$. With $T_i$ objective evaluations per intersection, full-search calibration costs $O\!\left(G_{\max}N_{\mathrm{pop}}\sum_{i=1}^{M}T_iN_iB_iC_f\right)$. The final fidelity value on $\mathcal{D}$ is cached, so its additional evaluation cost is $O(1)$ per candidate. Closed-loop simulation costs $O\!\left(G_{\max}N_{\mathrm{pop}}T_{\mathrm{sim}}(M\bar{B}C_f+C_{\mathrm{env}})\right)$ over the search. Simulation is typically the dominant cost in practice, followed by calibration.

\paragraph{Deployment.}
Scoring at most $B_i$ phases costs $O(B_iC_f)$ per intersection-level decision. The phase-scoring function is shared; each intersection stores its own phase biases.

\subsection{Limitations}

First, PPL-TSC depends on teacher rollouts for behavioral supervision, so teacher quality and state coverage influence the learned policy. Students may inherit undesirable teacher behavior or perform poorly in states underrepresented in the rollouts. Although the joint objective includes closed-loop traffic performance and students can outperform their teachers, this feedback does not eliminate the limitations of teacher-derived data. Future work could expand rollout coverage through student-driven data collection and further refine programmatic policies through closed-loop optimization.

Second, PPL-TSC learns students within a fixed PUF-based policy class. While this class works well for several teacher families, it may be insufficient for teachers whose behavior depends on richer state representations or more complex feature interactions. Future work could explore richer programmatic policy classes that preserve compactness while covering a broader range of teacher behaviors. The verifier checks generated functions on sampled traffic states and constructed test inputs. A function that passes these checks may still behave undesirably in unseen states or different road networks. The predefined semantic checks may not capture all aspects of reasonable control behavior. Applying them to different teachers requires considering the input features, action meanings, and their relationships, and may require adapting the checks accordingly.

Third, the program structure is discovered through LLM-guided generation and refinement, making the final policy dependent on the model’s ability to propose valid and useful candidate expressions. Although LLM-based generation achieves higher semantic-constraint pass rates than GA, MCTS, and random search under a matched time budget, the resulting policy structures may still be suboptimal. Future work could mitigate this dependence by integrating LLM-based synthesis with more systematic local search, stronger verification, and automated repair mechanisms.

\subsection{Broader Impacts}

PPL-TSC may improve the practical deployability of DRL-based traffic signal control by distilling neural controllers into compact programmatic policies. If used responsibly, such policies could help reduce traffic delay and improve control efficiency, with potential downstream benefits for fuel consumption and emissions. The compact programmatic representation may also make controller behavior easier to inspect, maintain, and adapt in resource-constrained deployments.

At the same time, traffic signal control is a safety-critical application, and results on standard simulators do not guarantee safe real-world deployment. Because PPL-TSC is teacher-guided, it may also inherit biases or failure modes from the teacher policy, which could affect different traffic participants unevenly. Real-world use therefore requires careful validation, human oversight, and additional safeguards before deployment.

\subsection{Licenses for Existing Assets}
\label{app:licenses}

This work uses existing simulators, traffic scenarios, controller designs, and LLM services only for research evaluation. We credit the corresponding original sources throughout the paper and summarize the asset usage and licensing information below.

\begin{itemize}[leftmargin=*, itemsep=2pt, topsep=2pt]
    \item \textbf{Traffic simulator.} We conduct closed-loop evaluation in CityFlow \citep{zhang2019cityflow}. CityFlow is publicly released under the Apache License 2.0 according to the CityFlow project repository and documentation.
    \item \textbf{Benchmark scenarios and traffic data.} The traffic scenarios used in our experiments are taken from the LibSignal benchmark \citep{mei2024libsignal} and related CityFlow-format traffic-signal-control benchmarks. LibSignal is released as an open-source research library.
    \item \textbf{Teacher controller designs.} We evaluate against CoLight \citep{wei2019colight}, FRAP \citep{zheng2019learning}, MPLight \citep{chen2020toward}, EcoLight \citep{chauhan2020ecolight}, and TinyLight \citep{xing2022tinylight}. These methods are credited through their original publications. Any publicly available reference implementations or benchmark implementations used to reproduce teacher behavior are used only under their associated repository licenses and terms; this manuscript does not redistribute third-party controller code.
    \item \textbf{LLM services.} PPL-TSC uses LLM backbones such as Gemini, Qwen, Claude, GPT, Grok, and MiniMax for candidate policy generation and repair through their official access channels. These services are used according to their respective provider terms of service. We do not redistribute model weights or proprietary service outputs as standalone assets.
    \item \textbf{Generated artifacts.} The tables, plots, learned programmatic policies, and analysis outputs reported in this paper are generated by the authors from the experimental pipeline. 
\end{itemize}

\begin{table}[tbp]
  \centering
  \caption{Scenario-level results for CoLight. Results are averaged over the five traffic flows in each scenario.}
  \label{tab:scenario_colight}
  \scriptsize
  \begin{threeparttable}
    \setlength{\tabcolsep}{3pt}
    \begin{tabular*}{\textwidth}{@{\extracolsep{\fill}}lccccc@{}}
      \toprule
      \multirow{2}{*}{\textbf{Scenario}} &
      \multicolumn{3}{c}{\textbf{ATT} $\downarrow$} &
      \multicolumn{2}{c}{\textbf{ABF} $\uparrow$} \\
      \cmidrule(lr){2-4}\cmidrule(lr){5-6}
      & \textbf{Teacher} & \textbf{Tailored} & \textbf{Unified}
      & \textbf{Tailored} & \textbf{Unified} \\
      \midrule
      Hangzhou1 & 96.00$\pm$5.20 & \textbf{94.31$\pm$3.45} & 98.63$\pm$4.76 & \textbf{0.66$\pm$0.06} & 0.60$\pm$0.06 \\
      Hangzhou2 & \textbf{122.47$\pm$4.06} & 123.52$\pm$2.71 & 127.24$\pm$3.66 & \textbf{0.68$\pm$0.08} & 0.62$\pm$0.09 \\
      Hangzhou3 & \textbf{80.08$\pm$1.87} & 82.39$\pm$1.15 & 84.65$\pm$2.37 & \textbf{0.73$\pm$0.04} & 0.71$\pm$0.04 \\
      Atlanta & \textbf{311.00$\pm$24.27} & 311.36$\pm$58.31 & 325.75$\pm$65.71 & \textbf{0.73$\pm$0.17} & 0.65$\pm$0.14 \\
      Jinan & 1109.23$\pm$311.54 & 937.99$\pm$433.65 & \textbf{358.95$\pm$27.54} & \textbf{0.60$\pm$0.31} & 0.15$\pm$0.04 \\
      Los Angeles & \textbf{846.30$\pm$16.21} & 853.04$\pm$13.30 & 852.89$\pm$12.48 & \textbf{0.95$\pm$0.05} & 0.90$\pm$0.09 \\
      Manhattan & 1668.92$\pm$90.51 & 607.53$\pm$104.25 & \textbf{502.55$\pm$85.26} & \textbf{0.62$\pm$0.05} & 0.56$\pm$0.05 \\
      \bottomrule
    \end{tabular*}
  \end{threeparttable}
\end{table}

\begin{table}[tbp]
  \centering
  \caption{Scenario-level results for FRAP. Results are averaged over the five traffic flows in each scenario.}
  \label{tab:scenario_frap}
  \scriptsize
  \begin{threeparttable}
    \setlength{\tabcolsep}{3pt}
    \begin{tabular*}{\textwidth}{@{\extracolsep{\fill}}lccccc@{}}
      \toprule
      \multirow{2}{*}{\textbf{Scenario}} &
      \multicolumn{3}{c}{\textbf{ATT} $\downarrow$} &
      \multicolumn{2}{c}{\textbf{ABF} $\uparrow$} \\
      \cmidrule(lr){2-4}\cmidrule(lr){5-6}
      & \textbf{Teacher} & \textbf{Tailored} & \textbf{Unified}
      & \textbf{Tailored} & \textbf{Unified} \\
      \midrule
      Hangzhou1 & 128.94$\pm$31.88 & 109.79$\pm$19.76 & \textbf{99.04$\pm$5.85} & \textbf{0.69$\pm$0.16} & 0.61$\pm$0.23 \\
      Hangzhou2 & 150.24$\pm$56.75 & 128.03$\pm$14.62 & \textbf{122.65$\pm$1.15} & \textbf{0.62$\pm$0.27} & 0.59$\pm$0.31 \\
      Hangzhou3 & 132.47$\pm$110.34 & 134.01$\pm$115.03 & \textbf{85.05$\pm$2.30} & \textbf{0.72$\pm$0.05} & 0.58$\pm$0.30 \\
      Atlanta & \textbf{258.77$\pm$25.82} & 333.29$\pm$31.27 & 364.11$\pm$13.07 & \textbf{0.75$\pm$0.10} & 0.66$\pm$0.11 \\
      Jinan & \textbf{307.61$\pm$15.17} & 312.17$\pm$2.58 & 316.78$\pm$1.92 & \textbf{0.62$\pm$0.02} & 0.59$\pm$0.03 \\
      Los Angeles & 714.95$\pm$74.80 & \textbf{442.73$\pm$69.32} & 466.25$\pm$74.95 & \textbf{0.47$\pm$0.14} & 0.45$\pm$0.10 \\
      Manhattan & 639.19$\pm$423.21 & \textbf{232.81$\pm$31.61} & 235.63$\pm$23.04 & \textbf{0.56$\pm$0.10} & 0.56$\pm$0.11 \\
      \bottomrule
    \end{tabular*}
  \end{threeparttable}
\end{table}

\begin{table}[tbp]
  \centering
  \caption{Scenario-level results for MPLight. Results are averaged over the five traffic flows in each scenario.}
  \label{tab:scenario_mplight}
  \scriptsize
  \begin{threeparttable}
    \setlength{\tabcolsep}{3pt}
    \begin{tabular*}{\textwidth}{@{\extracolsep{\fill}}lccccc@{}}
      \toprule
      \multirow{2}{*}{\textbf{Scenario}} &
      \multicolumn{3}{c}{\textbf{ATT} $\downarrow$} &
      \multicolumn{2}{c}{\textbf{ABF} $\uparrow$} \\
      \cmidrule(lr){2-4}\cmidrule(lr){5-6}
      & \textbf{Teacher} & \textbf{Tailored} & \textbf{Unified}
      & \textbf{Tailored} & \textbf{Unified} \\
      \midrule
      Hangzhou1 & 100.83$\pm$3.36 & \textbf{97.33$\pm$6.23} & 100.04$\pm$1.62 & \textbf{0.76$\pm$0.07} & 0.72$\pm$0.07 \\
      Hangzhou2 & \textbf{121.81$\pm$1.32} & 121.97$\pm$1.33 & 122.89$\pm$2.44 & \textbf{0.74$\pm$0.10} & 0.71$\pm$0.10 \\
      Hangzhou3 & 82.72$\pm$2.73 & \textbf{82.48$\pm$2.10} & 83.51$\pm$2.42 & \textbf{0.89$\pm$0.04} & 0.88$\pm$0.04 \\
      Atlanta & 249.45$\pm$15.03 & \textbf{236.32$\pm$5.22} & 241.55$\pm$4.61 & \textbf{0.79$\pm$0.11} & 0.71$\pm$0.15 \\
      Jinan & \textbf{297.07$\pm$7.04} & 307.66$\pm$4.84 & 314.77$\pm$6.00 & \textbf{0.72$\pm$0.10} & 0.68$\pm$0.08 \\
      Los Angeles & 588.16$\pm$99.49 & 579.86$\pm$61.50 & \textbf{544.45$\pm$166.22} & \textbf{0.58$\pm$0.04} & 0.56$\pm$0.15 \\
      Manhattan & \textbf{195.42$\pm$4.14} & 197.74$\pm$6.87 & 198.95$\pm$3.51 & \textbf{0.61$\pm$0.06} & 0.60$\pm$0.06 \\
      \bottomrule
    \end{tabular*}
  \end{threeparttable}
\end{table}

\begin{table}[tbp]
  \centering
  \caption{Scenario-level results for EcoLight (native) under its keep/switch action setting. Results are averaged over the five traffic flows in each scenario.}
  \label{tab:scenario_ecolight_native}
  \scriptsize
  \begin{threeparttable}
    \setlength{\tabcolsep}{3pt}
    \begin{tabular*}{\textwidth}{@{\extracolsep{\fill}}lccccc@{}}
      \toprule
      \multirow{2}{*}{\textbf{Scenario}} &
      \multicolumn{3}{c}{\textbf{ATT} $\downarrow$} &
      \multicolumn{2}{c}{\textbf{ABF} $\uparrow$} \\
      \cmidrule(lr){2-4}\cmidrule(lr){5-6}
      & \textbf{Teacher} & \textbf{Tailored} & \textbf{Unified}
      & \textbf{Tailored} & \textbf{Unified} \\
      \midrule
      Hangzhou1 & 205.88$\pm$37.67 & \textbf{183.19$\pm$25.05} & 194.82$\pm$15.38 & \textbf{0.81$\pm$0.05} & 0.77$\pm$0.05 \\
      Hangzhou2 & 135.03$\pm$0.54 & \textbf{134.63$\pm$2.11} & 135.92$\pm$2.79 & \textbf{0.91$\pm$0.04} & 0.90$\pm$0.04 \\
      Hangzhou3 & 98.05$\pm$1.77 & \textbf{96.31$\pm$1.24} & 97.23$\pm$1.52 & \textbf{0.93$\pm$0.02} & 0.92$\pm$0.03 \\
      Atlanta & \textbf{311.47$\pm$32.78} & 329.75$\pm$20.98 & 380.94$\pm$38.96 & \textbf{0.79$\pm$0.11} & 0.77$\pm$0.05 \\
      Jinan & \textbf{378.74$\pm$3.97} & 765.00$\pm$314.15 & 736.17$\pm$166.60 & \textbf{0.84$\pm$0.01} & 0.81$\pm$0.02 \\
      Los Angeles & \textbf{640.13$\pm$5.03} & 674.28$\pm$108.16 & 785.57$\pm$116.09 & 0.80$\pm$0.03 & \textbf{0.82$\pm$0.03} \\
      Manhattan & \textbf{1066.33$\pm$66.77} & 1325.00$\pm$16.36 & 1340.00$\pm$40.56 & \textbf{0.80$\pm$0.01} & 0.79$\pm$0.01 \\
      \bottomrule
    \end{tabular*}
  \end{threeparttable}
\end{table}

\begin{table}[tbp]
  \centering
  \caption{Scenario-level results for EcoLight (phase-selection). Results are averaged over the five traffic flows in each scenario.}
  \label{tab:scenario_ecolight}
  \scriptsize
  \begin{threeparttable}
    \setlength{\tabcolsep}{3pt}
    \begin{tabular*}{\textwidth}{@{\extracolsep{\fill}}lccccc@{}}
      \toprule
      \multirow{2}{*}{\textbf{Scenario}} &
      \multicolumn{3}{c}{\textbf{ATT} $\downarrow$} &
      \multicolumn{2}{c}{\textbf{ABF} $\uparrow$} \\
      \cmidrule(lr){2-4}\cmidrule(lr){5-6}
      & \textbf{Teacher} & \textbf{Tailored} & \textbf{Unified}
      & \textbf{Tailored} & \textbf{Unified} \\
      \midrule
      Hangzhou1 & 205.88$\pm$42.11 & 126.10$\pm$31.32 & \textbf{115.29$\pm$26.62} & \textbf{0.43$\pm$0.09} & 0.29$\pm$0.06 \\
      Hangzhou2 & 135.03$\pm$0.60 & 127.90$\pm$3.37 & \textbf{124.61$\pm$4.31} & \textbf{0.49$\pm$0.03} & 0.38$\pm$0.08 \\
      Hangzhou3 & 98.05$\pm$1.98 & \textbf{83.73$\pm$2.59} & 86.99$\pm$1.66 & \textbf{0.65$\pm$0.03} & 0.62$\pm$0.02 \\
      Atlanta & \textbf{311.47$\pm$36.65} & 358.34$\pm$55.18 & 381.22$\pm$11.42 & \textbf{0.56$\pm$0.05} & 0.48$\pm$0.07 \\
      Jinan & 378.74$\pm$4.44 & 324.74$\pm$9.79 & \textbf{320.34$\pm$3.63} & \textbf{0.39$\pm$0.02} & 0.28$\pm$0.03 \\
      Los Angeles & 640.13$\pm$5.62 & \textbf{485.39$\pm$35.87} & 530.55$\pm$98.75 & \textbf{0.44$\pm$0.05} & 0.41$\pm$0.06 \\
      Manhattan & 1066.33$\pm$74.65 & 249.80$\pm$29.01 & \textbf{241.49$\pm$33.82} & \textbf{0.42$\pm$0.03} & 0.36$\pm$0.02 \\
      \bottomrule
    \end{tabular*}
  \end{threeparttable}
\end{table}

\begin{table}[tbp]
  \centering
  \caption{Scenario-level results for TinyLight. Results are averaged over the five traffic flows in each scenario.}
  \label{tab:scenario_tinylight}
  \scriptsize
  \begin{threeparttable}
    \setlength{\tabcolsep}{3pt}
    \begin{tabular*}{\textwidth}{@{\extracolsep{\fill}}lccccc@{}}
      \toprule
      \multirow{2}{*}{\textbf{Scenario}} &
      \multicolumn{3}{c}{\textbf{ATT} $\downarrow$} &
      \multicolumn{2}{c}{\textbf{ABF} $\uparrow$} \\
      \cmidrule(lr){2-4}\cmidrule(lr){5-6}
      & \textbf{Teacher} & \textbf{Tailored} & \textbf{Unified}
      & \textbf{Tailored} & \textbf{Unified} \\
      \midrule
      Hangzhou1 & 98.76$\pm$5.81 & \textbf{97.11$\pm$6.78} & 105.11$\pm$15.31 & \textbf{0.67$\pm$0.07} & 0.57$\pm$0.06 \\
      Hangzhou2 & 121.84$\pm$1.20 & \textbf{119.83$\pm$0.78} & 123.16$\pm$5.12 & \textbf{0.66$\pm$0.08} & 0.57$\pm$0.09 \\
      Hangzhou3 & \textbf{81.17$\pm$1.22} & 81.53$\pm$1.11 & 83.86$\pm$3.85 & \textbf{0.82$\pm$0.08} & 0.70$\pm$0.09 \\
      Atlanta & \textbf{278.47$\pm$12.85} & 359.23$\pm$54.13 & 375.47$\pm$30.47 & \textbf{0.69$\pm$0.05} & 0.65$\pm$0.07 \\
      Jinan & \textbf{299.54$\pm$2.83} & 324.91$\pm$89.72 & 397.69$\pm$128.98 & 0.31$\pm$0.11 & \textbf{0.42$\pm$0.10} \\
      Los Angeles & \textbf{450.31$\pm$43.75} & 517.40$\pm$147.30 & 619.48$\pm$164.61 & \textbf{0.73$\pm$0.05} & 0.72$\pm$0.06 \\
      Manhattan & 762.03$\pm$532.47 & \textbf{562.91$\pm$664.60} & 1163.20$\pm$758.01 & 0.58$\pm$0.11 & \textbf{0.63$\pm$0.03} \\
      \bottomrule
    \end{tabular*}
  \end{threeparttable}
\end{table}

\subsubsection{Diagnosis of EcoLight (native) and TinyLight}
\label{app:teacher_specific_diagnosis}

We further investigate the weaker in-domain results obtained when PPL-TSC is applied to EcoLight (native) and TinyLight. We conduct controlled follow-up experiments over 35 tasks (seven scenarios and five traffic flows), changing only the suspected factor while keeping all other training and evaluation settings fixed. Table~\ref{tab:teacher_specific_diagnosis} reports average behavioral fidelity (ABF) and relative ATT improvement, where higher values are better.

\begin{table}[tbp]
  \centering
  \caption{Controlled diagnosis of EcoLight (native) and TinyLight. For EcoLight (native), Original inputs score phases separately using phase-local demand, whereas Aligned inputs include joint green- and red-side demand. The search budget is reported as (population size, number of generations).}
  \label{tab:teacher_specific_diagnosis}
  \scriptsize
  \setlength{\tabcolsep}{3pt}
  \begin{tabular}{@{}llcc@{}}
    \toprule
    \textbf{Teacher} & \textbf{Configuration} & \textbf{ABF $\uparrow$} & \textbf{$\Delta$ATT (\%) $\uparrow$} \\
    \midrule
    \multirow{2}{*}{EcoLight (native)}
      & Original inputs & 0.84 & $-18.05$ \\
      & Aligned inputs & \textbf{0.94} & \textbf{2.71} \\
    \midrule
    \multirow{2}{*}{TinyLight}
      & Search budget $(5,5)$ & 0.64 & $-8.43$ \\
      & Search budget $(10,10)$ & \textbf{0.71} & \textbf{1.51} \\
    \bottomrule
  \end{tabular}
\end{table}

\paragraph{EcoLight (native).}
EcoLight (native) makes keep/switch decisions using demand from both green and red movements, but the original student scores the current and next phases separately using only the demand each phase serves. An incorrect switch changes the current phase and the decisions available at subsequent steps, allowing errors to accumulate during closed-loop control. Using aligned inputs that provide joint green- and red-side demand, with all other settings unchanged, increases ABF from 0.84 to 0.94 and improves mean $\Delta$ATT from $-18.05\%$ to $+2.71\%$. These results suggest that the input mismatch is a major contributor to the original performance degradation.

\paragraph{TinyLight.}
TinyLight uses both phase-level and incoming-lane features, giving the student a broader set of inputs to search over. Increasing the population size and number of generations from $(5,5)$ to $(10,10)$, with all other settings unchanged, raises ABF from 0.64 to 0.71 and improves mean $\Delta$ATT from $-8.43\%$ to $+1.51\%$. These results suggest that the default budget limits search quality for TinyLight and that a larger budget can improve both fidelity and control performance.

\begin{table}[tbp]
  \centering
  \caption{Supplementary generalization results for CoLight under same-scenario, different-flow transfer. Cross-scenario results are not available in the current benchmark. Rows correspond to the source scenario.}
  \label{tab:transfer_colight}
  \scriptsize
  \begin{threeparttable}
    \begin{tabular}{@{}lcccc@{}}
      \toprule
      \textbf{Source Scenario} & \textbf{Teacher ATT $\downarrow$} & \textbf{PPL-CoLight ATT $\downarrow$} & \textbf{$\Delta$ATT (\%) $\uparrow$} & \textbf{ATT wins} \\
      \midrule
      Hangzhou1 & \textbf{96.74$\pm$4.70} & 99.36$\pm$2.41 & -2.96$\pm$5.72 & 4/20 \\
      Hangzhou2 & 122.85$\pm$3.26 & \textbf{122.36$\pm$1.07} & 0.33$\pm$2.61 & 6/20 \\
      Hangzhou3 & \textbf{81.04$\pm$1.28} & 86.72$\pm$1.18 & -7.04$\pm$2.17 & 0/20 \\
      Atlanta & \textbf{313.68$\pm$20.93} & 402.94$\pm$28.94 & -28.56$\pm$6.21 & 0/20 \\
      Jinan & 1114.97$\pm$254.41 & \textbf{751.00$\pm$474.67} & 35.48$\pm$32.05 & 15/20 \\
      Los Angeles & 846.71$\pm$16.73 & \textbf{636.83$\pm$188.32} & 24.57$\pm$23.06 & 16/20 \\
      Manhattan & 1660.76$\pm$89.71 & \textbf{202.42$\pm$8.44} & 87.79$\pm$0.66 & 20/20 \\
      \bottomrule
    \end{tabular}
  \end{threeparttable}
\end{table}

\begin{table}[tbp]
  \centering
  \caption{Supplementary generalization results for FRAP. Rows correspond to the source scenario.}
  \label{tab:transfer_frap}
  \scriptsize
  \begin{threeparttable}
    \begin{minipage}[t]{0.48\textwidth}
      \centering
      \begin{tabular}{@{}lcccc@{}}
        \toprule
        \multicolumn{5}{c}{\textbf{Same-Scenario, Different-Flow}} \\
        \midrule
        \textbf{Source Scenario} & \textbf{Teacher ATT $\downarrow$} & \textbf{PPL-FRAP ATT $\downarrow$} & \textbf{$\Delta$ATT (\%) $\uparrow$} & \textbf{ATT wins} \\
        \midrule
        Hangzhou1 & 126.50$\pm$27.48 & \textbf{100.03$\pm$2.89} & 18.10$\pm$13.19 & 20/20 \\
        Hangzhou2 & 142.02$\pm$40.37 & \textbf{122.95$\pm$1.49} & 8.99$\pm$15.76 & 18/20 \\
        Hangzhou3 & 134.77$\pm$101.22 & \textbf{86.61$\pm$1.07} & 12.54$\pm$30.97 & 7/20 \\
        Atlanta & \textbf{254.60$\pm$16.28} & 389.20$\pm$25.79 & -53.10$\pm$8.86 & 0/20 \\
        Jinan & \textbf{307.89$\pm$13.30} & 319.69$\pm$1.61 & -4.00$\pm$4.01 & 4/20 \\
        Los Angeles & 703.20$\pm$77.96 & \textbf{409.74$\pm$8.46} & 40.92$\pm$7.48 & 20/20 \\
        Manhattan & 675.21$\pm$396.23 & \textbf{196.57$\pm$4.76} & 59.44$\pm$20.08 & 20/20 \\
        \bottomrule
      \end{tabular}
    \end{minipage}\hfill
    \begin{minipage}[t]{0.48\textwidth}
      \centering
      \begin{tabular}{@{}lcccc@{}}
        \toprule
        \multicolumn{5}{c}{\textbf{Cross-Scenario}} \\
        \midrule
        \textbf{Source Scenario} & \textbf{Teacher ATT $\downarrow$} & \textbf{PPL-FRAP ATT $\downarrow$} & \textbf{$\Delta$ATT (\%) $\uparrow$} & \textbf{ATT wins} \\
        \midrule
        Hangzhou1 & 764.29$\pm$635.23 & \textbf{251.56$\pm$124.11} & 37.06$\pm$37.37 & 114/150 \\
        Hangzhou2 & 803.49$\pm$584.30 & \textbf{252.88$\pm$134.62} & 50.01$\pm$33.28 & 129/150 \\
        Hangzhou3 & 914.93$\pm$500.84 & \textbf{253.10$\pm$119.67} & 62.16$\pm$30.44 & 136/150 \\
        Atlanta & 1342.99$\pm$385.91 & \textbf{222.45$\pm$160.33} & 79.14$\pm$20.88 & 147/150 \\
        Jinan & 717.59$\pm$589.02 & \textbf{215.18$\pm$130.26} & 40.03$\pm$44.67 & 104/150 \\
        Los Angeles & 1128.05$\pm$548.72 & \textbf{199.71$\pm$111.34} & 66.97$\pm$36.88 & 138/150 \\
        Manhattan & 688.96$\pm$366.22 & \textbf{252.73$\pm$171.47} & 44.18$\pm$46.45 & 119/150 \\
        \bottomrule
      \end{tabular}
    \end{minipage}
  \end{threeparttable}
\end{table}

\begin{table}[tbp]
  \centering
  \caption{Supplementary generalization results for MPLight. Rows correspond to the source scenario.}
  \label{tab:transfer_mplight}
  \scriptsize
  \begin{threeparttable}
    \begin{minipage}[t]{0.48\textwidth}
      \centering
      \begin{tabular}{@{}lcccc@{}}
        \toprule
        \multicolumn{5}{c}{\textbf{Same-Scenario, Different-Flow}} \\
        \midrule
        \textbf{Source Scenario} & \textbf{Teacher ATT $\downarrow$} & \textbf{PPL-MPLight ATT $\downarrow$} & \textbf{$\Delta$ATT (\%) $\uparrow$} & \textbf{ATT wins} \\
        \midrule
        Hangzhou1 & 100.72$\pm$5.51 & \textbf{98.96$\pm$7.21} & 1.80$\pm$3.13 & 15/20 \\
        Hangzhou2 & 122.42$\pm$1.72 & \textbf{122.19$\pm$1.44} & 0.18$\pm$0.91 & 11/20 \\
        Hangzhou3 & \textbf{82.86$\pm$1.23} & 84.43$\pm$1.50 & -1.89$\pm$1.48 & 2/20 \\
        Atlanta & 251.50$\pm$17.14 & \textbf{243.17$\pm$11.01} & 2.87$\pm$7.81 & 10/20 \\
        Jinan & \textbf{297.09$\pm$6.17} & 310.32$\pm$1.13 & -4.50$\pm$2.12 & 1/20 \\
        Los Angeles & 591.77$\pm$55.95 & \textbf{537.51$\pm$156.13} & 8.35$\pm$27.90 & 16/20 \\
        Manhattan & \textbf{195.72$\pm$2.98} & 201.25$\pm$7.19 & -2.80$\pm$2.41 & 0/20 \\
        \bottomrule
      \end{tabular}
    \end{minipage}\hfill
    \begin{minipage}[t]{0.48\textwidth}
      \centering
      \begin{tabular}{@{}lcccc@{}}
        \toprule
        \multicolumn{5}{c}{\textbf{Cross-Scenario}} \\
        \midrule
        \textbf{Source Scenario} & \textbf{Teacher ATT $\downarrow$} & \textbf{PPL-MPLight ATT $\downarrow$} & \textbf{$\Delta$ATT (\%) $\uparrow$} & \textbf{ATT wins} \\
        \midrule
        Hangzhou1 & 792.01$\pm$647.69 & \textbf{256.48$\pm$155.39} & 40.67$\pm$34.82 & 115/150 \\
        Hangzhou2 & 798.34$\pm$661.12 & \textbf{233.04$\pm$126.91} & 44.34$\pm$33.91 & 124/150 \\
        Hangzhou3 & 843.40$\pm$531.29 & \textbf{240.28$\pm$120.13} & 58.53$\pm$27.53 & 150/150 \\
        Atlanta & 952.78$\pm$522.35 & \textbf{215.22$\pm$133.92} & 60.49$\pm$35.90 & 144/150 \\
        Jinan & 564.21$\pm$438.45 & \textbf{204.58$\pm$129.43} & 38.25$\pm$35.77 & 141/150 \\
        Los Angeles & 981.66$\pm$481.97 & \textbf{185.12$\pm$91.26} & 68.68$\pm$29.09 & 150/150 \\
        Manhattan & 288.98$\pm$240.39 & \textbf{231.98$\pm$161.99} & 11.29$\pm$20.26 & 122/150 \\
        \bottomrule
      \end{tabular}
    \end{minipage}
  \end{threeparttable}
\end{table}

\begin{table}[tbp]
  \centering
  \caption{Supplementary generalization results for EcoLight (native) under its keep/switch action setting. Rows correspond to the source scenario.}
  \label{tab:transfer_ecolight_keep_switch}
  \scriptsize
  \begin{threeparttable}
    \begin{minipage}[t]{0.48\textwidth}
      \centering
      \begin{tabular}{@{}lcccc@{}}
        \toprule
        \multicolumn{5}{c}{\textbf{Same-Scenario, Different-Flow}} \\
        \midrule
        \textbf{Source Scenario} & \textbf{Teacher ATT $\downarrow$} & \textbf{PPL-EcoLight ATT $\downarrow$} & \textbf{$\Delta$ATT (\%) $\uparrow$} & \textbf{ATT wins} \\
        \midrule
        Hangzhou1 & 209.20$\pm$35.75 & \textbf{98.73$\pm$3.15} & 51.52$\pm$7.63 & 20/20 \\
        Hangzhou2 & 134.42$\pm$1.02 & \textbf{122.49$\pm$1.27} & 8.87$\pm$1.15 & 20/20 \\
        Hangzhou3 & 97.86$\pm$3.08 & \textbf{86.62$\pm$1.21} & 11.40$\pm$2.95 & 20/20 \\
        Atlanta & \textbf{312.77$\pm$31.81} & 373.94$\pm$5.14 & -20.60$\pm$10.34 & 2/20 \\
        Jinan & 379.52$\pm$6.98 & \textbf{320.75$\pm$1.82} & 15.46$\pm$1.39 & 20/20 \\
        Los Angeles & 643.40$\pm$6.78 & \textbf{408.17$\pm$9.61} & 36.55$\pm$1.79 & 20/20 \\
        Manhattan & 1033.27$\pm$46.12 & \textbf{328.85$\pm$120.43} & 68.38$\pm$11.03 & 20/20 \\
        \bottomrule
      \end{tabular}
    \end{minipage}\hfill
    \begin{minipage}[t]{0.48\textwidth}
      \centering
      \begin{tabular}{@{}lcccc@{}}
        \toprule
        \multicolumn{5}{c}{\textbf{Cross-Scenario}} \\
        \midrule
        \textbf{Source Scenario} & \textbf{Teacher ATT $\downarrow$} & \textbf{PPL-EcoLight ATT $\downarrow$} & \textbf{$\Delta$ATT (\%) $\uparrow$} & \textbf{ATT wins} \\
        \midrule
        Hangzhou1 & 773.79$\pm$599.72 & \textbf{251.53$\pm$124.29} & 43.20$\pm$33.53 & 150/150 \\
        Hangzhou2 & 777.06$\pm$599.57 & \textbf{247.81$\pm$127.96} & 47.92$\pm$30.17 & 150/150 \\
        Hangzhou3 & 791.48$\pm$576.36 & \textbf{253.73$\pm$122.18} & 50.35$\pm$30.28 & 150/150 \\
        Atlanta & 750.53$\pm$543.70 & \textbf{205.91$\pm$121.93} & 60.80$\pm$25.66 & 150/150 \\
        Jinan & 537.76$\pm$534.79 & \textbf{228.62$\pm$160.51} & 35.39$\pm$30.98 & 139/150 \\
        Los Angeles & 957.33$\pm$626.34 & \textbf{200.58$\pm$111.70} & 60.95$\pm$36.34 & 125/150 \\
        Manhattan & \textbf{386.32$\pm$174.02} & 533.93$\pm$362.09 & -64.69$\pm$157.99 & 74/150 \\
        \bottomrule
      \end{tabular}
    \end{minipage}
  \end{threeparttable}
\end{table}

\begin{table}[tbp]
  \centering
  \caption{Supplementary generalization results for EcoLight (phase-selection). Rows correspond to the source scenario.}
  \label{tab:transfer_ecolight}
  \scriptsize
  \begin{threeparttable}
    \begin{minipage}[t]{0.48\textwidth}
      \centering
      \begin{tabular}{@{}lcccc@{}}
        \toprule
        \multicolumn{5}{c}{\textbf{Same-Scenario, Different-Flow}} \\
        \midrule
        \textbf{Source Scenario} & \textbf{Teacher ATT $\downarrow$} & \textbf{PPL-EcoLight ATT $\downarrow$} & \textbf{$\Delta$ATT (\%) $\uparrow$} & \textbf{ATT wins} \\
        \midrule
        Hangzhou1 & 209.20$\pm$35.75 & \textbf{107.25$\pm$12.43} & 47.05$\pm$11.14 & 20/20 \\
        Hangzhou2 & 134.42$\pm$1.02 & \textbf{123.58$\pm$1.66} & 8.06$\pm$1.43 & 20/20 \\
        Hangzhou3 & 97.86$\pm$3.08 & \textbf{87.07$\pm$1.80} & 10.92$\pm$3.59 & 20/20 \\
        Atlanta & \textbf{312.77$\pm$31.81} & 376.35$\pm$1.39 & -21.42$\pm$10.78 & 0/20 \\
        Jinan & 379.52$\pm$6.98 & \textbf{324.88$\pm$7.63} & 14.34$\pm$3.33 & 20/20 \\
        Los Angeles & 643.40$\pm$6.78 & \textbf{405.74$\pm$9.03} & 36.93$\pm$1.57 & 20/20 \\
        Manhattan & 1033.27$\pm$46.12 & \textbf{201.72$\pm$8.84} & 80.44$\pm$1.20 & 20/20 \\
        \bottomrule
      \end{tabular}
    \end{minipage}\hfill
    \begin{minipage}[t]{0.48\textwidth}
      \centering
      \begin{tabular}{@{}lcccc@{}}
        \toprule
        \multicolumn{5}{c}{\textbf{Cross-Scenario}} \\
        \midrule
        \textbf{Source Scenario} & \textbf{Teacher ATT $\downarrow$} & \textbf{PPL-EcoLight ATT $\downarrow$} & \textbf{$\Delta$ATT (\%) $\uparrow$} & \textbf{ATT wins} \\
        \midrule
        Hangzhou1 & 773.79$\pm$599.72 & \textbf{335.01$\pm$287.54} & 33.77$\pm$37.04 & 123/150 \\
        Hangzhou2 & 777.06$\pm$599.57 & \textbf{262.02$\pm$151.01} & 45.88$\pm$31.28 & 145/150 \\
        Hangzhou3 & 791.48$\pm$576.36 & \textbf{285.34$\pm$176.94} & 46.14$\pm$33.08 & 140/150 \\
        Atlanta & 750.53$\pm$543.70 & \textbf{205.71$\pm$120.23} & 60.67$\pm$25.67 & 150/150 \\
        Jinan & 537.76$\pm$534.79 & \textbf{220.34$\pm$140.70} & 36.58$\pm$30.25 & 149/150 \\
        Los Angeles & 957.33$\pm$626.34 & \textbf{200.30$\pm$111.34} & 61.00$\pm$36.27 & 125/150 \\
        Manhattan & 386.32$\pm$174.02 & \textbf{278.49$\pm$206.58} & 29.90$\pm$31.34 & 122/150 \\
        \bottomrule
      \end{tabular}
    \end{minipage}
  \end{threeparttable}
\end{table}

\begin{table}[tbp]
  \centering
  \caption{Supplementary generalization results for TinyLight under the same-scenario, different-flow transfer. Cross-scenario results are not available in the current benchmark. Rows correspond to the source scenario.}
  \label{tab:transfer_tinylight}
  \scriptsize
  \begin{threeparttable}
    \begin{tabular}{@{}lcccc@{}}
      \toprule
      \textbf{Source Scenario} & \textbf{Teacher ATT $\downarrow$} & \textbf{PPL-TinyLight ATT $\downarrow$} & \textbf{$\Delta$ATT (\%) $\uparrow$} & \textbf{ATT wins} \\
      \midrule
      Hangzhou1 & 99.85$\pm$7.36 & \textbf{95.99$\pm$3.09} & 3.56$\pm$4.48 & 16/20 \\
      Hangzhou2 & 121.68$\pm$1.44 & \textbf{119.59$\pm$1.27} & 1.71$\pm$0.96 & 20/20 \\
      Hangzhou3 & 81.76$\pm$1.10 & \textbf{81.19$\pm$1.30} & 0.70$\pm$1.30 & 15/20 \\
      Atlanta & \textbf{276.85$\pm$16.05} & 320.72$\pm$46.04 & -16.94$\pm$22.24 & 5/20 \\
      Jinan & 299.86$\pm$3.14 & \textbf{284.36$\pm$1.94} & 5.16$\pm$1.18 & 20/20 \\
      Los Angeles & \textbf{466.54$\pm$39.77} & 490.47$\pm$98.35 & -5.02$\pm$17.37 & 10/20 \\
      Manhattan & 611.26$\pm$485.68 & \textbf{373.68$\pm$330.27} & -20.54$\pm$126.79 & 12/20 \\
      \bottomrule
    \end{tabular}
  \end{threeparttable}
\end{table}

\subsection{Additional Experimental Details}
\label{app:additional_experimental_details}
\subsubsection{Experimental Setup}
\paragraph{Teacher Controllers.}
We use five representative DRL-based TSC methods as teachers. Most of them map intersection-level traffic observations to a phase-selection decision, while EcoLight uses a native keep/switch action interface. They differ substantially in how they represent traffic state and parameterize the control policy.

\textbf{CoLight} \citep{wei2019colight}. CoLight is a cooperative multi-intersection RL controller that uses graph attention to model interactions among neighboring intersections. For each intersection, the input includes the current phase and lane-level vehicle counts; these local features are further aggregated with features from adjacent intersections through an attention mechanism. The network outputs a value for each candidate phase, and the next phase is selected by taking the maximum-value action.

\textbf{FRAP} \citep{zheng2019learning}. FRAP models TSC from the perspective of phase competition and explicitly exploits the rotational symmetry of intersections. Its input consists of the current phase together with lane-link-level traffic features, which are reorganized into a phase-competition representation describing the demand served by competing phases. The output is a value for each feasible phase.

\textbf{MPLight} \citep{chen2020toward}. MPLight extends FRAP to large-scale road networks through parameter sharing across intersections and a pressure-based learning objective. Its input includes the current phase and lane-link-level pressure features, where pressure measures the imbalance between upstream and downstream traffic. Similar to FRAP, the network outputs a value for each candidate phase.

\textbf{EcoLight} \citep{chauhan2020ecolight}. EcoLight is a lightweight RL controller designed for budget- and infrastructure-constrained settings. It uses a compact low-dimensional state derived from the current phase and incoming-lane vehicle counts, and formulates control as a binary decision: keep the current phase or switch to the next phase. Its output is therefore a two-dimensional action-value vector over these two choices. In PPL-TSC, these two actions correspond to $\mathcal{P}_t^{(i)}=\{p_{t,\mathrm{cur}}^{(i)}, \mathrm{next}(p_{t,\mathrm{cur}}^{(i)})\}$. We report this native keep/switch setting as EcoLight (native), and additionally evaluate an EcoLight (phase-selection) variant in which the student directly scores all feasible phases.

\textbf{TinyLight} \citep{xing2022tinylight}. TinyLight is a resource-efficient DRL controller tailored for deployment on devices with extremely limited memory and computation. Its input includes compact phase-level and incoming-lane traffic statistics, such as vehicle count, queue length, waiting time, delay, pressure, and the current phase. The network outputs a value for each feasible phase, and the phase with the highest value is selected for execution.

\paragraph{Implementation details.}
All experiments are conducted on a machine with an Intel Core i7-10700K CPU, 64 GB RAM, and an NVIDIA GeForce RTX 4080 SUPER GPU with 16 GB VRAM. Teacher controllers are implemented in PyTorch and trained for 128 episodes. For programmatic policy search, PPL-TSC uses Gemini 3.1 Flash with a population size $N_{\mathrm{pop}}=5$ and a maximum of $G_{\max}=5$ generations, unless otherwise stated. For bias calibration, CMA-ES uses a population size of 20 and a maximum of 20 iterations.

\subsubsection{Scenario-Level Results}
\label{app:scenario_level_results}
This subsection provides the detailed scenario-level results corresponding to Table~\ref{tab:main_results}. Tables~\ref{tab:scenario_colight}--\ref{tab:scenario_tinylight} report results for CoLight, FRAP, MPLight, EcoLight (native), EcoLight (phase-selection), and TinyLight. For each teacher family, we report the mean and standard deviation over the five traffic flows in each scenario. Boldface indicates the best value within each scenario row, with lower values preferred for ATT and higher values preferred for ABF. These tables complement the aggregated summary in the main text by showing how the PPL-TSC (Tailored) and PPL-TSC (Unified) students behave across individual traffic scenarios.

\subsubsection{Cross-scenario Generalization Details}
\label{app:generalization_details}

This subsection provides supplementary transfer results corresponding to Table~\ref{tab:generalization}. Tables~\ref{tab:transfer_colight}--\ref{tab:transfer_tinylight} report transfer results for CoLight, FRAP, MPLight, EcoLight (native), EcoLight (phase-selection), and TinyLight. For each teacher family and each source scenario, we report average ATT under same-scenario, different-flow transfer; for teacher families with applicable settings, we additionally report cross-scenario transfer results. We focus on ATT here because the main goal of RQ3 is to assess whether the learned policy preserves traffic-control performance under distribution shift; compared with fidelity-oriented metrics, ATT provides a more direct measure of transfer effectiveness. Here, ``ATT wins'' counts the number of transfer test cases in which PPL-TSC improves upon or matches the teacher on ATT. For the same-scenario setting, each row aggregates over 20 transfer test cases; for the cross-scenario setting, each row aggregates over 150 transfer test cases when applicable. Cross-scenario transfer is not applicable to CoLight and TinyLight because their action dimensions vary across scenarios, whereas EcoLight, FRAP, and MPLight support both transfer settings.

\subsubsection{Additional Out-of-Distribution Evaluation}
\label{app:targeted_ood}
We further evaluate PPL-TSC under two targeted out-of-distribution (OOD) disturbances: (i) \textit{severe congestion}, where traffic demand increases to $1.5\times$ its original level during 40--60\% of an episode, and (ii) \textit{an accident}, simulated by closing all entry lanes of the highest-demand approach during the same period. All policies remain frozen, and each matched teacher--student pair uses the same network, traffic flow, random seed, disturbance, and simulation horizon. Each teacher interface is evaluated over seven scenarios, five traffic flows, and two disturbances, yielding 70 matched cases.

\begin{table}[htbp]
  \centering
  \caption{Performance under targeted OOD disturbances. ATT improvement is measured relative to the corresponding teacher; positive values indicate lower student ATT.}
  \label{tab:additional_ood}
  \scriptsize
  \begin{tabular}{lcc}
    \toprule
    \textbf{Teacher interface} & \textbf{$\Delta$ATT (\%)} & \textbf{ATT wins} \\
    \midrule
    CoLight & 0.66 & 35/70 \\
    EcoLight (phase-selection) & 2.66 & 48/70 \\
    EcoLight (native) & 1.07 & 30/70 \\
    FRAP & 1.26 & 38/70 \\
    MPLight & 1.19 & 40/70 \\
    TinyLight & -0.15 & 32/70 \\
    \midrule
    Overall & 1.12 & 223/420 \\
    \bottomrule
  \end{tabular}
\end{table}

Under the two targeted disturbances, five of six teacher interfaces achieve positive mean improvements, with an overall improvement of 1.12\% and 223/420 wins. These results show that PPL-TSC remains broadly competitive with its teachers under severe congestion and accident-induced lane closures.

\subsubsection{Comparison with Distill-Adapted Programmatic Baselines}
\label{app:distill_baselines}
We compare PPL-TSC with three representative programmatic policy-search methods: $\pi$-Light \citep{gu2024pi}, GPLight+ \citep{liao2025gplight+}, and SymLight \citep{liao2025symlight}. Because their original formulations learn through environment interaction rather than teacher rollouts, we adapt them into $\pi$-Light-Distill, GPLight+-Distill, and SymLight-Distill for a controlled teacher-guided comparison. These adapted baselines use the same MPLight teacher rollouts, phase interface, search budget, simulator settings, and evaluation protocol as PPL-TSC; SymLight-Distill is reimplemented from its paper because public code is unavailable. No method uses target-scenario search, calibration, or teacher data after transfer.

Each scenario contains five traffic flows, with one policy learned per source flow. For the 42 directed cross-scenario pairs, five source policies are evaluated on five target flows, yielding 25 evaluations per pair and 1,050 per method. For the seven same-scenario pairs, each policy is evaluated on the other four flows, yielding 20 evaluations per pair and 140 per method. This produces 1,190 evaluations per method and 4,760 evaluations in total.

\begin{table}[htbp]
  \centering
  \caption{Aggregate comparison with distill-adapted programmatic baselines. Cross-scenario ATT is reported as mean $\pm$ SD, with the median in parentheses, over the 1,050 off-diagonal evaluations. Best-pair counts and mean ranks use pair-level median ATT over all 49 source--target pairs.}
  \label{tab:distill_baseline_summary}
  \scriptsize
  \begin{tabular*}{\textwidth}{@{\extracolsep{\fill}}lccc@{}}
    \toprule
    \textbf{Method} & \textbf{Cross-scenario ATT (s) $\downarrow$} & \textbf{Best pairs $\uparrow$} & \textbf{Mean rank $\downarrow$} \\
    \midrule
    \textbf{PPL-TSC} & \textbf{232.51 $\pm$ 154.20 (217.01)} & \textbf{24/49} & \textbf{1.94} \\
    $\pi$-Light-Distill & 305.99 $\pm$ 274.68 (230.98) & 3/49 & 3.21 \\
    GPLight+-Distill & 301.32 $\pm$ 286.78 (234.88) & 11/49 & 2.54 \\
    SymLight-Distill & 275.23 $\pm$ 249.27 (220.53) & 11/49 & 2.31 \\
    \bottomrule
  \end{tabular*}
\end{table}

PPL-TSC achieves the lowest cross-scenario mean and median ATT, obtains the best result on 24 of 49 source--target pairs, and has the lowest overall mean rank. Table~\ref{tab:distill_transfer_matrix} provides the complete transfer matrix. Each cell reports median ATT in the order PPL-TSC / $\pi$-Light-Distill / GPLight+-Distill / SymLight-Distill; off-diagonal and diagonal cells aggregate 25 cross-scenario and 20 held-out same-scenario evaluations, respectively.

\begin{table*}[htbp]
  \centering
  \caption{All-to-all transfer comparison with distill-adapted programmatic baselines. Each cell reports median ATT as PPL-TSC / $\pi$-Light-Distill / GPLight+-Distill / SymLight-Distill; boldface denotes the lowest value.}
  \label{tab:distill_transfer_matrix}
  \scriptsize
  \resizebox{\textwidth}{!}{%
  \begin{tabular}{lccccccc}
    \toprule
    \textbf{Source / Target} & \textbf{Atlanta} & \textbf{Hangzhou1} & \textbf{Hangzhou2} & \textbf{Hangzhou3} & \textbf{Jinan} & \textbf{Los Angeles} & \textbf{Manhattan} \\
    \midrule
    Atlanta & 239.05 / 236.23 / 236.24 / \textbf{235.12} & 98.57 / 100.18 / 100.94 / \textbf{98.20} & \textbf{121.98} / 122.96 / 122.89 / 122.32 & \textbf{85.06} / 85.96 / 86.08 / 85.35 & \textbf{310.48} / 972.14 / 451.72 / 327.69 & \textbf{446.30} / 494.71 / 473.96 / 523.34 & \textbf{216.61} / 551.18 / 284.36 / 217.70 \\
    Hangzhou1 & \textbf{238.91} / 307.04 / 246.65 / 262.26 & 98.97 / \textbf{97.50} / 100.13 / 101.19 & \textbf{121.98} / 122.89 / 122.34 / 122.11 & 85.09 / \textbf{84.01} / 85.24 / 84.90 & \textbf{311.48} / 342.63 / 330.78 / 325.05 & \textbf{458.47} / 763.85 / 604.76 / 700.44 & \textbf{217.37} / 245.91 / 256.26 / 229.73 \\
    Hangzhou2 & \textbf{239.32} / 245.45 / 243.67 / 248.47 & 99.13 / 96.18 / \textbf{95.20} / 96.18 & 122.21 / \textbf{122.07} / 122.25 / 122.19 & 85.33 / 84.66 / \textbf{84.35} / 84.66 & 316.33 / 331.47 / \textbf{308.03} / 313.59 & 454.33 / 487.48 / \textbf{453.35} / 490.67 & 220.93 / 230.42 / \textbf{216.49} / 222.03 \\
    Hangzhou3 & 246.89 / 245.41 / \textbf{244.26} / 308.25 & 95.70 / 100.12 / \textbf{93.81} / 95.97 & 122.14 / 122.49 / \textbf{121.99} / 122.03 & 84.43 / 85.07 / 83.86 / \textbf{83.56} & 316.08 / 321.09 / 309.96 / \textbf{303.85} & \textbf{470.13} / 484.35 / 542.14 / 763.90 & \textbf{210.03} / 221.39 / 224.13 / 218.08 \\
    Jinan & 238.90 / 238.67 / \textbf{238.57} / 239.04 & 99.03 / 98.66 / 98.94 / \textbf{98.31} & \textbf{121.98} / 122.32 / 122.62 / 122.10 & 85.09 / 85.51 / 85.51 / \textbf{84.82} & 310.53 / 310.86 / 309.31 / \textbf{309.12} & \textbf{445.59} / 460.05 / 483.16 / 472.98 & 216.53 / 218.87 / 214.51 / \textbf{213.95} \\
    Los Angeles & \textbf{240.33} / 243.00 / 341.65 / 241.37 & \textbf{96.87} / 102.88 / 106.42 / 100.10 & \textbf{122.00} / 123.42 / 125.53 / 123.30 & \textbf{85.32} / 86.08 / 91.05 / 86.08 & \textbf{313.00} / 941.63 / 339.10 / 355.16 & 466.45 / 490.53 / \textbf{461.39} / 466.50 & 217.82 / 226.46 / \textbf{215.96} / 221.52 \\
    Manhattan & \textbf{238.69} / 262.08 / 244.86 / 241.96 & 98.85 / 98.66 / 97.90 / \textbf{96.87} & 122.22 / 122.89 / 122.32 / \textbf{122.00} & \textbf{84.96} / 86.08 / 85.33 / 85.36 & \textbf{310.95} / 340.81 / 328.71 / 325.27 & \textbf{453.28} / 666.24 / 463.61 / 482.43 & 213.92 / 216.89 / 209.93 / \textbf{207.39} \\
    \bottomrule
  \end{tabular}%
  }
\end{table*}

\subsubsection{Readability Examples Across Teacher Families}
\label{app:multi_teacher_readability}

Table~\ref{tab:multi_teacher_readability} extends the MPLight case study in Section~\ref{sec:readability} to all five teacher families and both EcoLight interfaces. The table shows the phase-scoring component $f_p=f(\mathbf{x}_{p,t}^{(i)})$; ATT and ABF are measured using the complete policy $U_p=f_p+\beta_p^{(i)}$. The EcoLight (native) example uses the aligned inputs described in Appendix~\ref{app:teacher_specific_diagnosis}.

\begin{table}[tbp]
  \centering
  \caption{Representative learned rules and traffic-control interpretations across teacher families. ATT is measured in seconds. Each row corresponds to one scenario--flow task.}
  \label{tab:multi_teacher_readability}
  \scriptsize
  \setlength{\tabcolsep}{3pt}
  \renewcommand{\arraystretch}{1.2}
  \renewcommand{\tabularxcolumn}[1]{m{#1}}
  \begin{tabularx}{\textwidth}{@{}>{\raggedright\arraybackslash}m{1.95cm}>{\raggedright\arraybackslash}m{3.4cm}>{\raggedright\arraybackslash}Xrrr@{}}
    \toprule
    \multirow{2}{*}{\textbf{Teacher / task}} & \multirow{2}{*}{\textbf{Phase-scoring component}} & \multirow{2}{*}{\textbf{Traffic-control interpretation}} & \multirow{2}{*}{\textbf{ABF $\uparrow$}} & \multicolumn{2}{c}{\textbf{ATT (s) $\downarrow$}} \\
    \cmidrule(lr){5-6}
    & & & & Teacher & Student \\
    \midrule
    \textbf{CoLight}\newline Atlanta / flow 3
      & $f_p=V_p+0.2I_pV_p^2$ & Prioritizes demand; a quadratic bonus retains a busy active phase. & 0.94 & 339.20 & 317.90 \\
    \addlinespace[8pt]
    \textbf{EcoLight}\newline (phase-selection)\newline Hangzhou2 / flow 2
      & $\begin{aligned}f_p&=0.7V_p+0.3V_p^{\max}\\&\quad+0.15I_p\end{aligned}$ & Combines total and peak demand with an active-phase retention bonus. & 0.94 & 135.93 & 134.13 \\
    \addlinespace[8pt]
    \textbf{EcoLight}\newline (native, aligned)\newline Hangzhou2 / flow 2
      & $\begin{aligned}f_{\mathrm{keep}}&=\tfrac{g}{\sqrt{g+1}}\\[3pt]f_{\mathrm{switch}}&=\tfrac{r\log(1+r)}{g+1}\end{aligned}$ & Keeps green for served demand; switches as red-side demand grows relative to green-side demand. & 0.99 & 135.93 & 135.05 \\
    \addlinespace[8pt]
    \textbf{FRAP}\newline Atlanta / flow 1
      & $f_p=\dfrac{\sum_{d\in\mathcal{R}_p}n_d}{17}+0.2I_p$ & Scores aggregate vehicle demand linearly and adds a fixed bonus for the active phase. & 0.89 & 255.03 & 245.01 \\
    \addlinespace[8pt]
    \textbf{MPLight}\newline Hangzhou3 / flow 4
      & $\begin{aligned}f_p&=P_p^{\Sigma}+P_p^{\max}\\&\quad+0.15I_p\end{aligned}$ & Combines aggregate and bottleneck pressure with a retention bonus, following MaxPressure~\citep{varaiya2013max}. & 0.93 & 86.01 & 84.04 \\
    \addlinespace[8pt]
    \textbf{TinyLight}\newline Hangzhou3 / flow 1
      & $\begin{aligned}f_p&=P_p^{\Sigma}+0.5\log(1+W_p)\\&\quad+0.2N_p+0.5I_p\end{aligned}$ & Prioritizes pressure, queues, and waiting time, with a retention bonus. The logarithm gives additional waiting time progressively less weight. & 0.86 & 80.77 & 80.63 \\
    \bottomrule
  \end{tabularx}
\end{table}

Here, $I_p$ equals 1 when phase $p$ is active and 0 otherwise. $V_p$ and $V_p^{\max}$ denote total and peak vehicle demand, respectively; $N_p$ and $W_p$ denote queued vehicle count and accumulated waiting time. For brevity, the decision-step index $t$ and intersection index $i$ are omitted from the table expressions. The pressure terms follow Fig.~\ref{fig:student_policy_evo}. For the set of directions $\mathcal{R}_p$ served by phase $p$, $P_p^{\Sigma}=\sum_{d\in\mathcal{R}_p}\hat{p}_{d,t}$ and $P_p^{\max}=\max_{d\in\mathcal{R}_p}\hat{p}_{d,t}$, where $\hat{p}_{d,t}$ is normalized directional pressure. Before normalization, directional pressure is the upstream vehicle count minus the downstream vehicle count. For EcoLight (native, aligned), $g$ and $r$ denote green- and red-side demand, respectively. For FRAP, $n_d$ denotes directional vehicle demand, and the constant 17 is the 95th percentile of \texttt{vehicles\_raw} in the source data.

\subsubsection{LLM Configuration, Cost, and Randomness}
\label{app:llm_cost_randomness}
The main experiments use Gemini 3.1 Flash with temperature 0.8; top-$p$ and maximum output length use the API defaults, and individual responses contain 407--779 tokens. To evaluate random-seed sensitivity, we conduct five searches for each of five teachers and seven scenarios, giving 175 searches. Within each of the 35 teacher--scenario combinations, we compute the coefficient of variation (CV; standard deviation divided by mean) of the final joint score across the five seeds and then average the CV across combinations. The remaining quantities in Table~\ref{tab:llm_cost_randomness} are averaged over all 175 searches.

\begin{table}[htbp]
  \centering
  \caption{LLM random-seed sensitivity and discovery cost.}
  \label{tab:llm_cost_randomness}
  \scriptsize
  \begin{tabular}{lc}
    \toprule
    \textbf{Metric} & \textbf{Mean} \\
    \midrule
    Final joint-score CV & 2.08\% \\
    Token usage per search & 67.1K \\
    Generation cost per search & USD 0.0353 \\
    LLM API latency per search & 104.2 s \\
    \bottomrule
  \end{tabular}
\end{table}

The low average final-score CV indicates limited sensitivity to sampling randomness under this protocol. LLM use is confined to policy discovery: reproducing a search requires the same model and API configuration and may still vary because of provider-side sampling, whereas exported policies execute deterministically as standalone Python code and require no LLM during inference or deployment. Token cost and API latency are provider- and time-dependent and are reported only for the evaluated configuration.

\begin{figure*}[!h]
  \centering
  \includegraphics[width=0.9\textwidth]{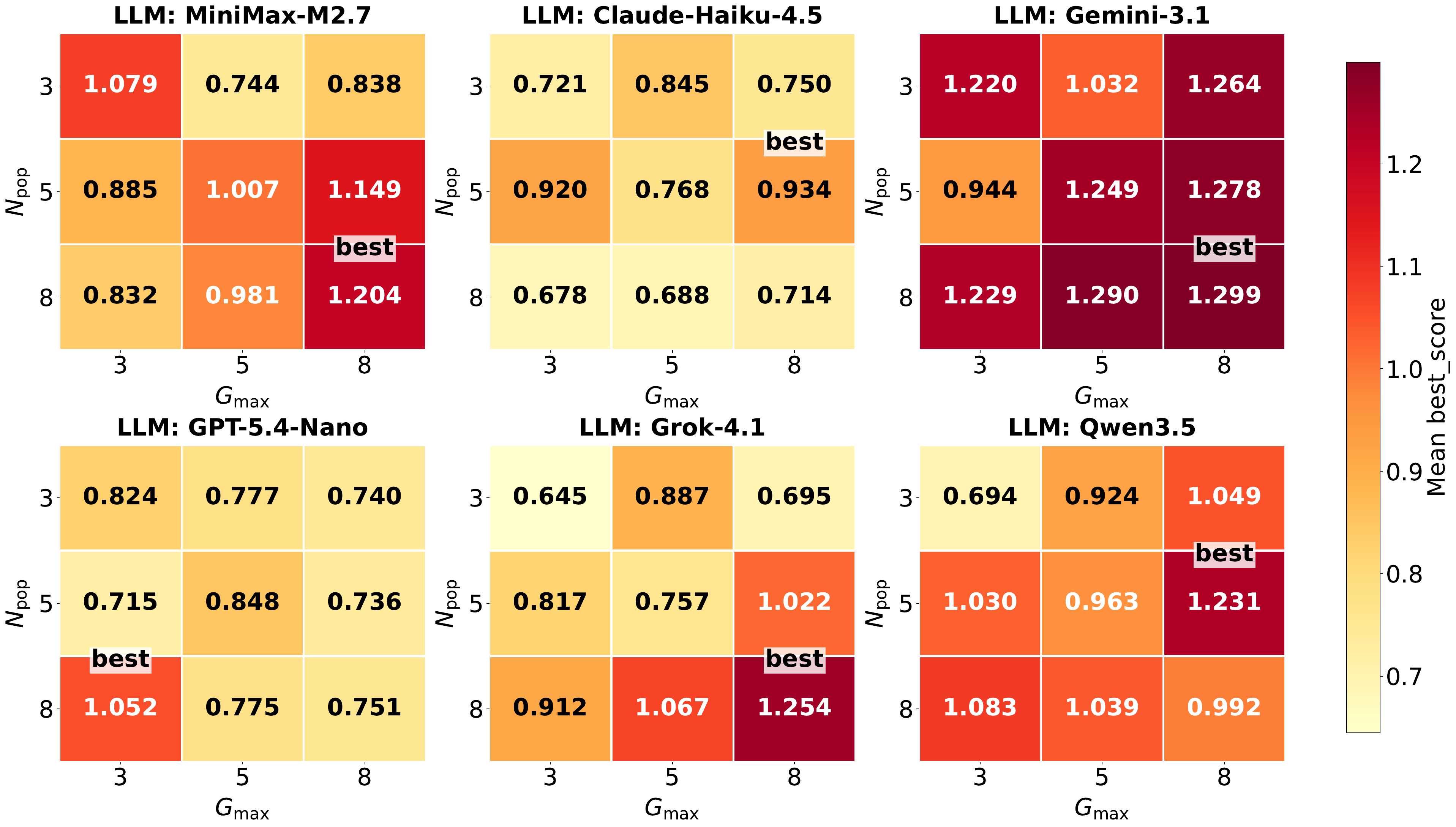}
  \caption{Parameter sensitivity on MPLight/Los Angeles/flow0. Each heatmap shows the final best score under one LLM backbone, where the score is computed by the policy evaluation objective in Eq.~\eqref{eq:evaluation}. Rows correspond to $N_{\mathrm{pop}}$ and columns to $G_{\max}$.}
  \label{fig:param_sensitivity}
\end{figure*}

\subsubsection{Parameter Analysis}
We analyze the sensitivity of PPL-TSC to three configuration factors: the LLM backbone, the population size $N_{\mathrm{pop}}$, and the number of generations $G_{\max}$. Fig.~\ref{fig:param_sensitivity} visualizes the 54 runs on MPLight/Los Angeles/flow0 as per-backbone heatmaps of the final best score, where the score is the policy evaluation objective defined in Eq.~\eqref{eq:evaluation}.

Two observations are clear from the figure. First, the LLM backbone substantially affects the absolute search quality. Gemini 3.1 Flash is the strongest overall, with all nine settings above 0.94 and a peak score of 1.299 at $(8,8)$. Grok-4.1 and Qwen3.5 form a second tier, reaching 1.254 and 1.231, respectively. By contrast, Claude-Haiku-4.5 and GPT-5.4-Nano are weaker, with best scores of only 0.934 and 1.052.

Second, larger generation budgets are usually more helpful than simply enlarging the population. For five of the six backbones, the best cell lies at $G_{\max}=8$; the only exception is GPT-5.4-Nano, whose best score appears at $(8,3)$. The preferred population size is less stable: the best setting uses $N_{\mathrm{pop}}=8$ for MiniMax-M2.7, Gemini 3.1 Flash, GPT-5.4-Nano, and Grok-4.1, but $N_{\mathrm{pop}}=5$ for Claude-Haiku-4.5 and Qwen3.5. Overall, $(5,5)$ provides a conservative default, while larger-budget settings such as $(5,8)$ and $(8,8)$ offer stronger search quality when additional runtime is available.

\clearpage
\section*{NeurIPS Paper Checklist}

\begin{enumerate}

\item {\bf Claims}
    \item[] Question: Do the main claims made in the abstract and introduction accurately reflect the paper's contributions and scope?
    \item[] Answer: \answerYes{}
    \item[] Justification: The abstract and introduction accurately summarize the paper's core contributions: PPL-TSC, the phase utility function representation, and the generate--review--calibrate--evaluate framework. Their claims on transparency, teacher fidelity, transfer generalization, and computational efficiency are consistent with the experimental results reported in the main paper and appendix.

\item {\bf Limitations}
    \item[] Question: Does the paper discuss the limitations of the work performed by the authors?
    \item[] Answer: \answerYes{}
    \item[] Justification: The paper includes a dedicated Limitations subsection in the appendix that discusses key constraints of the current approach, including teacher dependence and LLM-guided search dependence. It also outlines practical implications and future directions to mitigate these limitations.

\item {\bf Theory assumptions and proofs}
    \item[] Question: For each theoretical result, does the paper provide the full set of assumptions and a complete (and correct) proof?
    \item[] Answer: \answerNA{}
    \item[] Justification: The paper does not present formal theorems, lemmas, or proof-based theoretical claims. The main contributions are methodological and empirical.

    \item {\bf Experimental result reproducibility}
    \item[] Question: Does the paper fully disclose all the information needed to reproduce the main experimental results of the paper to the extent that it affects the main claims and/or conclusions of the paper (regardless of whether the code and data are provided or not)?
    \item[] Answer: \answerYes{}
    \item[] Justification: The paper and appendix describe the method pipeline, evaluation protocol, teacher settings, key hyperparameters, and implementation choices. The public repository at \url{https://github.com/ForrestXie9/PPL-TSC} provides implementation code, traffic networks and flows, configuration templates, and commands for teacher training, trajectory collection, policy learning, and evaluation. Teacher weights, trajectories, and learned policies can be generated using these commands; policy discovery requires user-configured LLM API access.
    \item[] Guidelines:
    \begin{itemize}
        \item The answer \answerNA{} means that the paper does not include experiments.
        \item If the paper includes experiments, a \answerNo{} answer to this question will not be perceived well by the reviewers: Making the paper reproducible is important, regardless of whether the code and data are provided or not.
        \item If the contribution is a dataset and\slash or model, the authors should describe the steps taken to make their results reproducible or verifiable. 
        \item Depending on the contribution, reproducibility can be accomplished in various ways. For example, if the contribution is a novel architecture, describing the architecture fully might suffice, or if the contribution is a specific model and empirical evaluation, it may be necessary to either make it possible for others to replicate the model with the same dataset, or provide access to the model. In general, releasing code and data is often one good way to accomplish this, but reproducibility can also be provided via detailed instructions for how to replicate the results, access to a hosted model (e.g., in the case of a large language model), releasing of a model checkpoint, or other means that are appropriate to the research performed.
        \item While NeurIPS does not require releasing code, the conference does require all submissions to provide some reasonable avenue for reproducibility, which may depend on the nature of the contribution. For example
        \begin{enumerate}
            \item If the contribution is primarily a new algorithm, the paper should make it clear how to reproduce that algorithm.
            \item If the contribution is primarily a new model architecture, the paper should describe the architecture clearly and fully.
            \item If the contribution is a new model (e.g., a large language model), then there should either be a way to access this model for reproducing the results or a way to reproduce the model (e.g., with an open-source dataset or instructions for how to construct the dataset).
            \item We recognize that reproducibility may be tricky in some cases, in which case authors are welcome to describe the particular way they provide for reproducibility. In the case of closed-source models, it may be that access to the model is limited in some way (e.g., to registered users), but it should be possible for other researchers to have some path to reproducing or verifying the results.
        \end{enumerate}
    \end{itemize}

\item {\bf Open access to data and code}
    \item[] Question: Does the paper provide open access to the data and code, with sufficient instructions to faithfully reproduce the main experimental results, as described in supplemental material?
    \item[] Answer: \answerYes{}
    \item[] Justification: The code and traffic networks and flows are publicly available at \url{https://github.com/ForrestXie9/PPL-TSC}. The README specifies the required environment and commands for installation, teacher training, trajectory collection, policy learning, and same-scenario and cross-scenario evaluation. Pretrained weights, collected trajectories, and generated policies are not bundled and must be produced using the provided pipeline. LLM-based policy discovery requires API access and may vary with provider-side sampling.

\item {\bf Experimental setting/details}
    \item[] Question: Does the paper specify all the training and test details (e.g., data splits, hyperparameters, how they were chosen, type of optimizer) necessary to understand the results?
    \item[] Answer: \answerYes{}
    \item[] Justification: The manuscript specifies the teacher families, scenario setup, optimization choices, and key hyperparameters for policy search and calibration, with additional details in the appendix. The test settings and evaluation metrics used in the main comparisons are also explicitly described.

\item {\bf Experiment statistical significance}
    \item[] Question: Does the paper report error bars suitably and correctly defined or other appropriate information about the statistical significance of the experiments?
    \item[] Answer: \answerYes{}
    \item[] Justification: The paper reports variability using mean and standard deviation (shown as $\pm$ values in tables and summarized in appendix tables). The aggregation protocol for these statistics is described alongside the reported results.

\item {\bf Experiments compute resources}
    \item[] Question: For each experiment, does the paper provide sufficient information on the computer resources (type of compute workers, memory, time of execution) needed to reproduce the experiments?
    \item[] Answer: \answerNo{}
    \item[] Justification: Appendix~\ref{app:additional_experimental_details} specifies the CPU, GPU, memory, training episodes, and search settings. Appendix~\ref{app:llm_cost_randomness} reports LLM API latency, but per-experiment execution times and the total compute budget are not fully reported.
    
    
\item {\bf Code of ethics}
    \item[] Question: Does the research conducted in the paper conform, in every respect, with the NeurIPS Code of Ethics \url{https://neurips.cc/public/EthicsGuidelines}?
    \item[] Answer: \answerYes{}
    \item[] Justification: We have reviewed the NeurIPS Code of Ethics and are not aware of any aspect of this work that violates it. The study uses simulated traffic environments and does not involve human-subject experimentation.

\item {\bf Broader impacts}
    \item[] Question: Does the paper discuss both potential positive societal impacts and negative societal impacts of the work performed?
    \item[] Answer: \answerYes{}
    \item[] Justification: The manuscript includes a dedicated Broader Impacts subsection in the appendix. That discussion covers both potential benefits (e.g., improved traffic efficiency and reduced congestion-related emissions) and potential risks, together with cautionary deployment considerations.
    
\item {\bf Safeguards}
    \item[] Question: Does the paper describe safeguards that have been put in place for responsible release of data or models that have a high risk for misuse (e.g., pre-trained language models, image generators, or scraped datasets)?
    \item[] Answer: \answerNA{}
    \item[] Justification: This work does not release high-risk generative models or scraped datasets that would require specialized misuse safeguards. The method targets traffic-signal control in simulation settings.

\item {\bf Licenses for existing assets}
    \item[] Question: Are the creators or original owners of assets (e.g., code, data, models), used in the paper, properly credited and are the license and terms of use explicitly mentioned and properly respected?
    \item[] Answer: \answerYes{}
    \item[] Justification: The original simulator, benchmark, and teacher methods are credited through their original publications, with asset usage described in Appendix~\ref{app:licenses}. The public code repository provides third-party attribution in \texttt{THIRD\_PARTY\_NOTICES.md} and includes license files for the bundled simulator and dependencies. We respect the applicable licenses and terms of use for third-party code, data, and LLM services.

\item {\bf New assets}
    \item[] Question: Are new assets introduced in the paper well documented and is the documentation provided alongside the assets?
    \item[] Answer: \answerYes{}
    \item[] Justification: The released implementation is documented in the repository README at \url{https://github.com/ForrestXie9/PPL-TSC}, including installation requirements, supported teachers and student variants, configuration templates, execution commands, and the scope of included assets. Third-party attribution and licensing notes are provided in \texttt{THIRD\_PARTY\_NOTICES.md}.

\item {\bf Crowdsourcing and research with human subjects}
    \item[] Question: For crowdsourcing experiments and research with human subjects, does the paper include the full text of instructions given to participants and screenshots, if applicable, as well as details about compensation (if any)? 
    \item[] Answer: \answerNA{}
    \item[] Justification: The study does not involve crowdsourcing or human-subject data collection. All experiments are conducted in simulation environments.

\item {\bf Institutional review board (IRB) approvals or equivalent for research with human subjects}
    \item[] Question: Does the paper describe potential risks incurred by study participants, whether such risks were disclosed to the subjects, and whether Institutional Review Board (IRB) approvals (or an equivalent approval/review based on the requirements of your country or institution) were obtained?
    \item[] Answer: \answerNA{}
    \item[] Justification: No human participants are involved in this work, so IRB approval and participant-risk disclosure are not applicable.

\item {\bf Declaration of LLM usage}
    \item[] Question: Does the paper describe the usage of LLMs if it is an important, original, or non-standard component of the core methods in this research? Note that if the LLM is used only for writing, editing, or formatting purposes and does \emph{not} impact the core methodology, scientific rigor, or originality of the research, declaration is not required.
    \item[] Answer: \answerYes{}
    \item[] Justification: LLM usage is a core part of the method: the model is used for candidate policy generation and iterative repair in the generate--review--calibrate--evaluate pipeline. The paper describes this role in both the abstract/method section and appendix prompt details.

\end{enumerate}
\end{document}